\documentclass[oldversion]{aa}

\usepackage{amsmath}
\usepackage{natbib}
\usepackage{epsfig}
\usepackage{txfonts}

\newcommand{\id}{{\,\rm d}}

\newcommand{\beq}{\begin{equation}}   %

\newcommand{\eeq}{\end{equation}}   %

\newcommand{\beqa}{\begin{eqnarray}}   %

\newcommand{\eeqa}{\end{eqnarray}}   %

\newcommand{\beal}{\begin{align}}

\newcommand{\enal}{\end{align}}

\newcommand{\bspl}{\begin{split}}

\newcommand{\espl}{\end{split}}

\newcommand{\bsub}{\begin{subequations}}

\newcommand{\esub}{\end{subequations}}

\newcommand{\bmulti}{\begin{multline}}   %

\newcommand{\beqm}{\begin{mathletters}}   %

\newcommand{\eeqm}{\end{mathletters}}   %

\newcommand{\Abst}[1]{\,#1}

\newcommand{\kB}{k_{\rm B}}

\newcommand{\me}{m_{\rm e}}

\newcommand{\Te}{T_{\rm e}}

\newcommand{\Tg}{T_{\gamma}}

\newcommand{\The}{\theta_{\rm e}}

\newcommand{\sigT}{\sigma_{\rm T}}

\newcommand{\pd}{\partial}

\newcommand{\pAb}[2]{\frac{\displaystyle\pd #1}{\displaystyle\pd #2}}

\newcommand{\pot}[2]{#1 \times 10^{#2}}

\newcommand{\xD}{{{x_{\rm D}}}}

\newcommand{\zi}{{z_{\rm i}}}

\usepackage{color}
\newcommand{\change}[1]{{#1}}

\begin{document}

\titlerunning{Cosmological Hydrogen Recombination: influence of resonance and electron scattering}
  
  
\title{Cosmological hydrogen recombination: influence of \\resonance and electron scattering}
 
 \author{J. Chluba\inst{1, 2} \and R.A. Sunyaev\inst{1,3}}
\authorrunning{Chluba \and Sunyaev}

\institute{Max-Planck-Institut f\"ur Astrophysik, Karl-Schwarzschild-Str. 1,
85741 Garching bei M\"unchen, Germany
\and
Canadian Institute for Theoretical Astrophysics, 60 St. George Street,
Toronto, ON M5S 3H8, Canada 
\and 
Space Research Institute, Russian Academy of Sciences, Profsoyuznaya 84/32,
117997 Moscow, Russia
}

\offprints{J. Chluba, 
\\ \email{jchluba@mpa-garching.mpg.de}
}

\date{Received / Accepted}

\abstract{In this paper we consider the effects of {\it resonance} and {\it electron scattering} on the escape of Lyman $\alpha$ photons during cosmological hydrogen recombination. We pay particular attention to the influence of {\it atomic recoil}, {\it Doppler boosting} and {\it Doppler broadening} using a Fokker-Planck approximation of the redistribution function describing the scattering of photons on the Lyman $\alpha$ resonance of moving hydrogen atoms.
We extend the computations of our recent paper on the influence of the 3d/3s-1s two-photon channels on the dynamics of hydrogen recombination, simultaneously including the full {\it time-dependence} of the problem, the {\it thermodynamic corrections factor}, leading to a frequency-dependent asymmetry between the emission and absorption profile, and the {\it quantum-mechanical corrections} related to the two-photon nature of the 3d/3s-1s emission and absorption process on the exact {\it shape} of the Lyman $\alpha$ emission profile.
We show here that due to the redistribution of photons over frequency hydrogen recombination is sped up by $\Delta N_{\rm e}/ N_{\rm e}\sim -0.6\%$ at $z\sim 900$. For the CMB temperature and polarization power spectra this results in $|\Delta C_l/C_l |\sim 0.5\%-1\%$ at $l \gtrsim 1500$, and therefore will be important for the analysis of future CMB data in the context of the {\sc Planck} Surveyor, {\sc Spt} and {\sc Act}.
The main contribution to this correction is coming from the atomic recoil effect ($\Delta N_{\rm e}/ N_{\rm e}\sim -1.2\%$ at $z\sim 900$), while Doppler boosting and Doppler broadening partially cancel this correction, again slowing hydrogen recombination down by $\Delta N_{\rm e}/ N_{\rm e}\sim 0.6\%$  at $z\sim 900$.
The influence of electron scattering close to the maximum of the Thomson visibility function at $z\sim 1100$ can be neglected.
We also give the cumulative results when in addition including the time-dependent correction, the thermodynamic factor and the correct shape of the emission profile. This amounts in $\Delta N_{\rm e}/ N_{\rm e}\sim -1.8\%$ at $z\sim 1160$ and $|\Delta C_l/C_l |\sim 1\%-3\%$ at $l \gtrsim 1500$.
}

\maketitle

\section{Introduction}
\label{sec:Intro}
Motivated by the great experimental prospects with the {\sc Planck} surveyor, {\sc Spt} and {\sc Act} several independent groups \citep[e.g.][]{Dubrovich2005, Chluba2006, Kholu2006, Jose2006, Switzer2007I, Wong2007} have investigated details in the physics of cosmological recombination and their impact on the theoretical predictions for the cosmic microwave background
(CMB) temperature and polarization power spectra.
The declared goal for our theoretical understanding of the ionization history is the $\sim 0.1\%$ accuracy level \citep[e.g. see][]{Hu1995, Seljak2003} close to the maximum of the Thomson visibility function \citep{Sunyaev1970} at $z\sim 1100$ \citep[e.g. see][for a more detailed overview of the different previously neglected physical processes that are important at this level of accuracy]{Sunyaev2008,Fendt2008}.

This paper is a continuation of our recent work on cosmological recombination, in which we studied the effects of 3d-1s and 3s-1s two-photon processes on the dynamics of hydrogen recombination \citep{Chluba2009}.  
Here we now wish to give the results for the changes in the Lyman $\alpha$ escape probability and free electron fraction when  in addition accounting for the effects {\it partial frequency redistribution} related to resonance scattering of moving neutral atoms and {\it electron scattering} during this epoch. In our previous work we neglected this aspect of the problem, although in the standard textbook formulation based on a Fokker-Planck expansion of the frequency redistribution function \citep{Rybicki2006} we obtained these results already some time ago.  Here we explain the main results of these computations which we also partly used elsewhere \citep{Jose2008, Chluba2008a}, and also refine our computations including the 3d-1s and 3s-1s two-photon corrections.
%

%
It is well known \citep[e.g. see][]{RybickiDell94} that for the conditions in our Universe (practically no collisions) the frequency redistribution function for photons scattering off moving atoms is given by the so called type-II redistribution as defined in \citet{Hummer1962}. 
The main physical processes which are accounted for in the Fokker-Planck expansion of this frequency redistribution function are due to (i) {\it atomic recoil},  (ii) {\it Doppler boosting}, and (iii) {\it Doppler broadening}.
\change{All three physical processes are also well-known in connection with the Kompaneets equation which describes the repeated scattering of photons by free electrons.}
Atomic recoil leads to a systematic drift of photons towards lower frequencies after each resonance scattering. This allows some additional photons to escape from the Lyman $\alpha$ resonance and thereby speeds hydrogen recombination up, as already demonstrated earlier by \citet{Grachev2008}. 
We found very similar results for this process some time ago \citep[e.g. see footnote 10 in][]{Chluba2008b}, which here we shall present in detail and also refine including additional corrections simultaneously.

However, in the analysis of \citet{Grachev2008} the effect due to (ii) and (iii) were not taken into account. Like atomic recoil Doppler boosting leads to a systematic motion of photons, but this time towards higher frequencies. Therefore it is expected to slow recombination down.
In contrast to this Doppler broadening can lead to both an increase or a decrease in the escape probability depending on where the photon initially is emitted.
As we explain here, if the photons are initially emitted in the vicinity of the Doppler core line diffusion helps to bring some of them towards the red wing, before they actually {\it die} (mainly due to two-photon absorption to the third shell). Similarly, for photons emitted on the blue side of the resonance line broadening allows some finite number of them to transverse the Doppler core. In the no line scattering approximation\footnote{In this approximation only true line emission and line absorption and redshifting of photons are included in the computation. The redistribution of photons over frequency is neglected.} this would not be possible, so that in both case the escape fraction is increased.
In contrast to this, for photons emitted on the red side of the resonance the effect of Doppler broadening decreases the escape fraction, since even up to $\sim 100$ Doppler width below the line center a significant fraction of the photons still returns close to the Doppler core, where they die efficiently.
As we show here, the combination of Doppler boosting and Doppler broadening in total leads to an additional decrease in the escape probability as compared to the no line scattering approximation.

For the expected correction due to electron scattering very similar arguments apply. However, there are some important differences: (i) electron scattering is expected to become less important at lower redshifts, since the free electron fraction decreases with time; (ii) in contrast to resonance scattering for Lyman $\alpha$ photons the electron scattering cross section is {\it achromatic}; and (iii) due to the smaller mass of the electron the recoil effect is $\sim 2000$ larger.
\change{Nevertheless, it turns out that during hydrogen recombination electron scattering can be neglected in the analysis of future CMB data. This is because of its much smaller cross section in comparison with line scattering and the decreasing number density of free electrons (see Sect.~\ref{sec:escatt}).}

We would like to mention that while this paper was in preparation another investigation of this problem was carried out by \citet{Hirata2009}. The results obtained in their work seem to be in good agreement with those presented here.

\section{Additions to the kinetic equation for the photons in the vicinity of the
  Lyman $\alpha$ resonance}
\label{sec:kin_eq}
Here we give the additional terms for the photon radiative transfer equation which are necessary to describe the effect of resonance and electron scattering in the Lyman $\alpha$ escape problem during cosmological hydrogen recombination.
We will use the same notation as in \citet{Chluba2008b} and \citet{Chluba2009}, also introducing the dimensionless frequency variable $x=\nu/(1+z)$ and photon distribution, $\tilde{N}_{x}=N_{x}/(1+z)^3=N_\nu/(1+z)^2$, with
$N_{\nu}=I_{\nu}/h\nu$, where $I_{\nu}$ is the {\it physical} specific
intensity of the ambient radiation field.
The photon occupation number then is $n_{\nu}=I_{\nu}/2h\nu^3$.
%
\change{Note that with this choice of variables the redshifting of photons due to the Hubble expansion is automatically taken into account in $x$ \citep[for more details see][]{Chluba2008b}. }

It is clear that Lyman-$\alpha$ line and electron scattering (both
including the Doppler-broadening, recoil and induced scatterings) only lead to
the {\it redistribution} of photons over frequency, but do not change the
total number of photons in each event. Also a blackbody spectrum with
$\Tg=\Te$ should not be altered by these processes.
Within the Fokker-Planck formulation of the corresponding processes these requirements are directly fulfilled.

\begin{figure*}
\centering 
\includegraphics[width=0.9\columnwidth]
{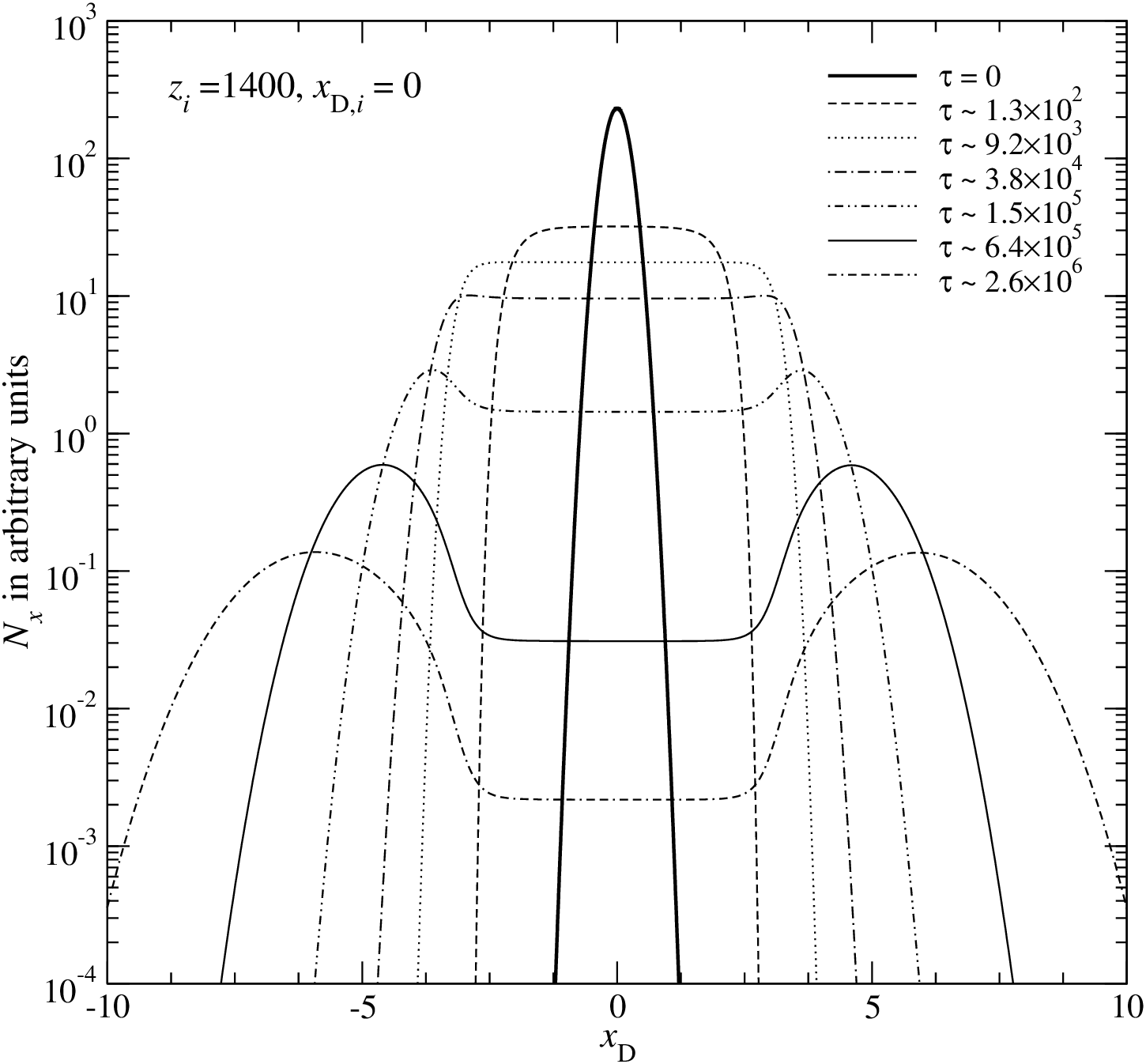}
\hspace{5mm}
\includegraphics[width=0.885\columnwidth]
{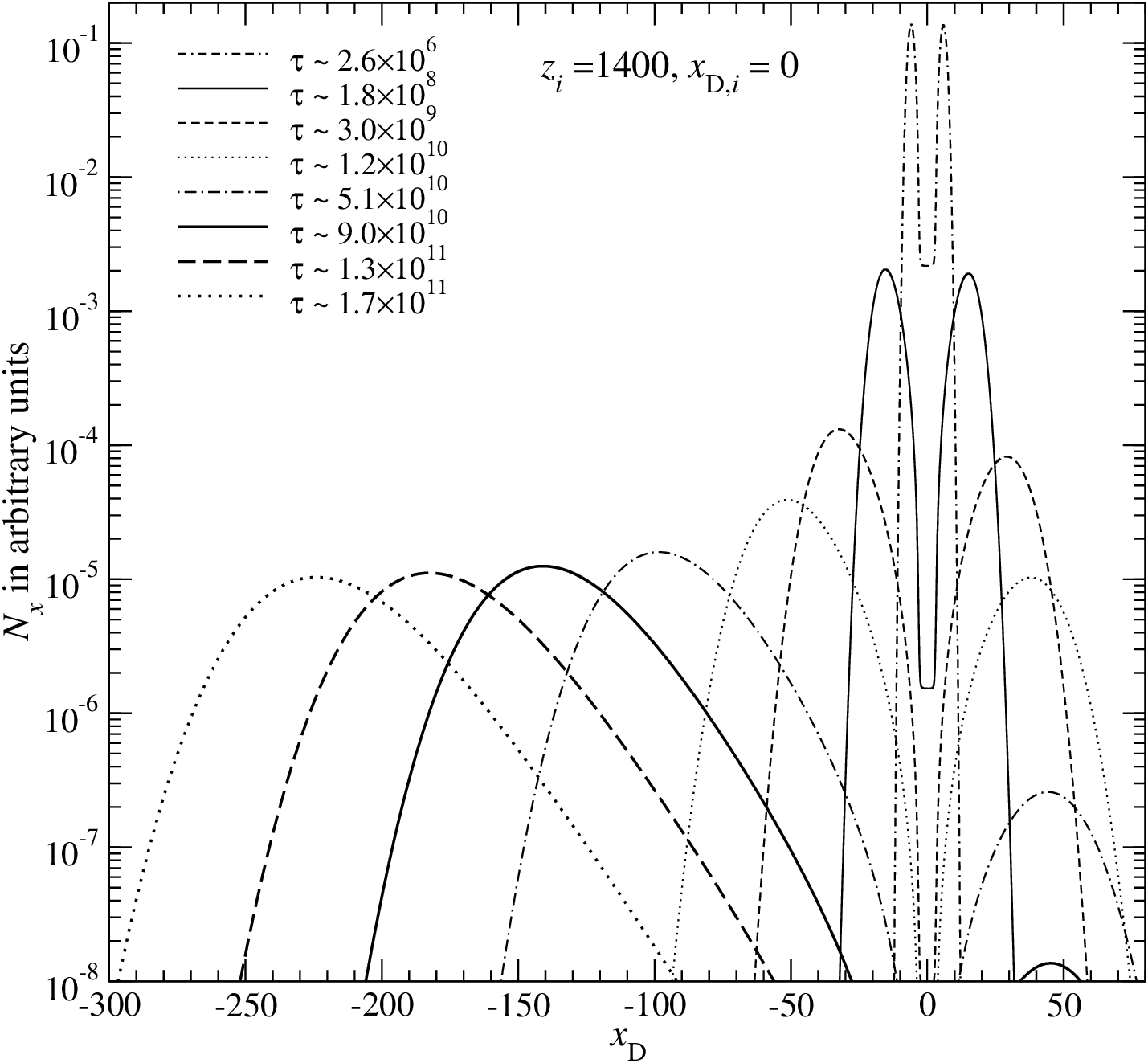}
\caption
{Time evolution of the photon distribution for single (narrow-line) injection
at the line center. The death probability for a 3-shell hydrogen atom was used
and electron scattering has been neglected. We use the time-variable $\tau=\int c \,\sigma_{\rm r} N_{\rm H}\id t$.}
\label{fig:line_1400_xD_0}
\end{figure*}

\subsection{Lyman-$\alpha$ resonance scattering}
\label{app:Ly-a-scatt}
The contribution to the collision term due to redistribution of photon by
resonance scattering off moving atoms can be written as \citep[e.g.
see][]{Rybicki2006}
\beal
\label{app:coll_N_nu_r}
\left.\mathcal{C}[N_{\nu}]\right|_{\rm r}
&=\int\mathcal{R}(\nu,\nu')N_{\nu'}(1+n_{\nu}) \id\nu'
\nonumber\\
&\qquad\qquad\qquad
-\int \mathcal{R}(\nu',\nu) N_{\nu}(1+n_{\nu'}) \id\nu'
\Abst{,} 
\end{align}
where $\mathcal{R}(\nu, \nu')$ is the frequency redistribution function for
the scattering atom, which for conditions in the Universe (practically no
electron or proton collisions!) is purely due to the Doppler effect
\citep[type-II redistribution as defined in][]{Hummer1962}. 

%
As shown in \citet{Rybicki2006}, within a Fokker-Planck formulation for the
case of Doppler redistribution Eq.~\eqref{app:coll_N_nu_r} can be cast into
the form
\beal
\label{app:coll_N_nu_r_appr}
\left.\mathcal{C}[N_{\nu}]\right|_{\rm r}
&\approx
p^{1\gamma}_{\rm sc}\,\sigma_{\rm r} N_{1\rm s}\frac{\Delta\nu_{\rm D}^2}{2}
\nonumber\\
&\,\,\quad\times
\pAb{}{\nu}\,\nu^2\phi_{\rm V}(\nu)
\left[\pAb{}{\nu} \frac{N_{\nu}}{\nu^2}
+\frac{h}{k\Te}\frac{N_{\nu}}{\nu^2}\left(1+\frac{c^2 N_\nu}{2\nu^2}\right)\right],
\end{align}
where $\sigma_{\rm r}=\frac{h\nu_{21}}{4\pi}\,\frac{B_{12}}{\Delta\nu_{\rm D}}$ denotes the resonant scattering cross section and $\Delta\nu_{\rm D}$ the Doppler width of the Lyman $\alpha$ resonance.
\change{The first term in brackets ($\propto\partial_\nu n_\nu$) describes the combined effect of Doppler boosting (it is of the order $\sim V^2/c^2$, where $V$ is the velocity of the atom) and Doppler broadening ($\sim V/c$), while the second term ($\propto n_\nu[1+n_\nu]$) accounts for atomic recoil ($\sim h\nu/m_{\rm H}c^2$) and stimulated scatterings.}
 Following \citet{RybickiDell94} we have used the diffusion coefficient
 $D\propto\phi_{\rm V}(\nu)$, where $\phi_{\rm V}(\nu)$ is the normal Voigt-profile. 
 We will neglect corrections due to non-resonant contributions \citep[e.g. see][]{Lee2005} in the scattering cross section, which would lead to a different frequency dependence far away from the resonance (e.g. Rayleigh scattering in the distant red wing \citep{Jackson}).

 It is important to note that Eq.~\eqref{app:coll_N_nu_r_appr} simultaneously includes the effects of {\it line diffusion}, {\it atomic recoil}\footnote{This terms was first introduced by \citet{Basko1978a, Basko1981}} and {\it stimulated scattering}\footnote{For the escape of Lyman $\alpha$ photons during hydrogen recombination this term is not important.}.  In this formulation it therefore
 preserves a Planckian photon distribution $N^{\rm Pl}_\nu$ with $\Tg\equiv\Te$.
 This can be easily verified when realizing that
 $\frac{\partial}{\partial_\nu} \frac{N^{\rm Pl}_\nu}{\nu^2}\equiv -
 \frac{h}{k\Tg}\frac{N^{\rm Pl}_{\nu}}{\nu^2}\left(1+\frac{c^2 N^{\rm
       Pl}_\nu}{2\nu^2}\right)$.
 Also one can easily verify that in the Fokker-Planck formulation $\int
 \left.\mathcal{C}[N_{\nu}]\right|_{\rm r} \id \nu=0$.

In Eq.~\eqref{app:coll_N_nu_r_appr} we also took into account the fact that 
not every scattering leads to the reappearance of the photon, since per
scattering the fraction $1-p_{\rm sc}^{1\gamma}$ of photons disappear in other channels,
i.e to higher levels and the continuum.
Here $p^{1\gamma}_{\rm sc}$ is the {\it single scattering albedo} which in
our formulation is equivalent to the one photon emission probability $p^{1\gamma}_{\rm em}$.
However, since $p^{1\gamma}_{\rm sc}$ is always very close to unity \citep[e.g. see][]{Chluba2009},
one could also neglect this detail here.

The corresponding term in the variables $x$ and $\tilde{N}_x$ then reads
\beal
\label{app:coll_N_x_r_appr}
\left.\mathcal{C}[\tilde{N}_x]\right|_{\rm r}
&\approx
p^{1\gamma}_{\rm em}\,\sigma_{\rm r} N_{1\rm s}\frac{\Delta\nu_{\rm D}^2}{2\eta^2}
\nonumber\\
&\,\,\quad\times
\pAb{}{x}\,x^2\phi_{\rm V}(\nu)
\left[\pAb{}{x} \frac{\tilde{N}_{x}}{x^2}
+\xi\,\frac{\tilde{N}_{x}}{x^2}\left(1+\frac{c^2 \tilde{N}_{x}}{2 x^2}\right)\right] 
\end{align}
where we have made the substitutions $\eta=(1+z)$, $\xi=\frac{h\eta}{k\Te}$ and $\nu=x[1+z]$.
Note that $\frac{\Delta\nu_{\rm D}^2}{2\eta^2}=x_{21}^2 \frac{k\Te}{m_{\rm H}c^2}$, where $x_{21}=\nu_{21}/[1+z]$ and $m_{\rm H}$ is the mass of the hydrogen atom.
\change{This term has to be added to the radiative transfer equation which includes the effect of line emission and absorption and can be found in \citet{Chluba2008b} for the normal '$1+1$' photon formulation of the problem and in \citet{Chluba2009} for the two-photon formulation.}

\subsection{Electron scattering}
\label{sec:escatt}
The contribution to the collision term due to scattering off free,
non-relativistic electrons can be described with the Kompaneets-equation.
Due to the similarity with Eq.~\eqref{app:coll_N_nu_r_appr} \citep[see
also][]{Rybicki2006} it is straightforward to obtain the corresponding terms
for our set of variables:
\beal
\label{app:coll_N_x_C}
\left.\mathcal{C}[\tilde{N}_x]\right|_{\rm C}
&\!\!=\!\sigT\,N_{\rm e}\,\The\,
\pAb{}{x}\,x^4\left[\pAb{}{x} \frac{\tilde{N}_{x}}{x^2}
+\xi\,\frac{\tilde{N}_{x}}{x^2}\left(1+\frac{c^2 \tilde{N}_{x}}{2 x^2}\right)\right] 
\Abst{,} 
\end{align}
where $\sigT\approx \pot{6.65}{-25}\,\text{cm}^2$ is the Thomson cross section
and $\The=\kB\Te/\me c^2$. 
Again one can clearly see that the electron scattering term preserves a
Planckian photon spectrum for $\Tg\equiv\Te$, and that $\int
\left.\mathcal{C}[N_{\nu}]\right|_{\rm r} \id \nu=0$.

\subsubsection{Relative importance of electron scattering}
Since the line-profile $\phi_{\rm V}$ is a strong function of frequency, resonance
scattering is most important close to the Lyman $\alpha$ line center, while in
the very distant wings electron scattering is expected to dominate.
Comparing the diffusion coefficients in frequency space for resonant and electron scattering
\bsub
\label{eq:DEDL}
\beal
\frac{\sigT\,N_{\rm e}\,\The\,x^2}
{p^{1\gamma}_{\rm sc}\sigma _{\rm r} N_{1\rm s}\frac{\Delta\nu_{\rm D}^2}{2\eta^2}\,\phi_{\rm V}(\nu)}
&\!\approx\!\pot{1.9}{-10}\,\frac{N_{\rm e}}{N_{1\rm s}}\frac{(1+z)^{1/2}}{\phi_{\rm V}(\nu)}
\\
&\!\!\!\!\stackrel{\rm wings}{\approx} 
\pot{2.1}{-8}\,(1+z)\,\frac{N_{\rm e}}{N_{1\rm s}}\,x_{\rm D}^{2}, 
\end{align}
\esub
\change{shows that at redshift $z\sim 1400$ (where $N_{\rm e}/N_{1\rm s} \sim 4$) in the line center resonance scattering is $\sim \pot{2.0}{7}$ times more important than electron scattering, and only at $|\xD|\gtrsim 100$ Doppler width electron scattering is able to
compete with line scattering. 
%
}

Due to the changes in $N_{\rm e}/N_{\rm 1s}$ the ratio \eqref{eq:DEDL} is a
strong function of redshift. However, electron scattering is expected to
influence the evolution of photons close to the line center significantly only
at redshifts $z\gtrsim 2500$, i.e. well before the main epoch of hydrogen
recombination.
Therefore one expects that electron scattering has a small impact on the
development of the photons close to the center of the Lyman-$\alpha$
transition and hence on the escape probability during hydrogen recombination.

\section{Illustrative time-dependent solutions for different initial photon distributions}
\label{sec:illustr}
In this Sect. we illustrate the main physical effects related to resonance scattering and electron scattering.
For this we numerically solved the radiative transfer equation injecting a
single narrow-line at different distances from the line center. For the computations we include the frequency redistribution of photons, redshifting and real absorption using the normal '$1+1$' photon picture \citep[see][]{Chluba2008b}. 
We neglect the effects due to two-photon corrections here.
Furthermore, we shall assume that the solution for the electron number 
density and the 1s-population are given by the output of the {\sc Recfast} code \citep{SeagerRecfast1999}.  
A few words about the PDE-solver can be found in the Appendix~\ref{app:comp_details}.

\begin{figure*}
\centering 
\includegraphics[width=0.90\columnwidth]
{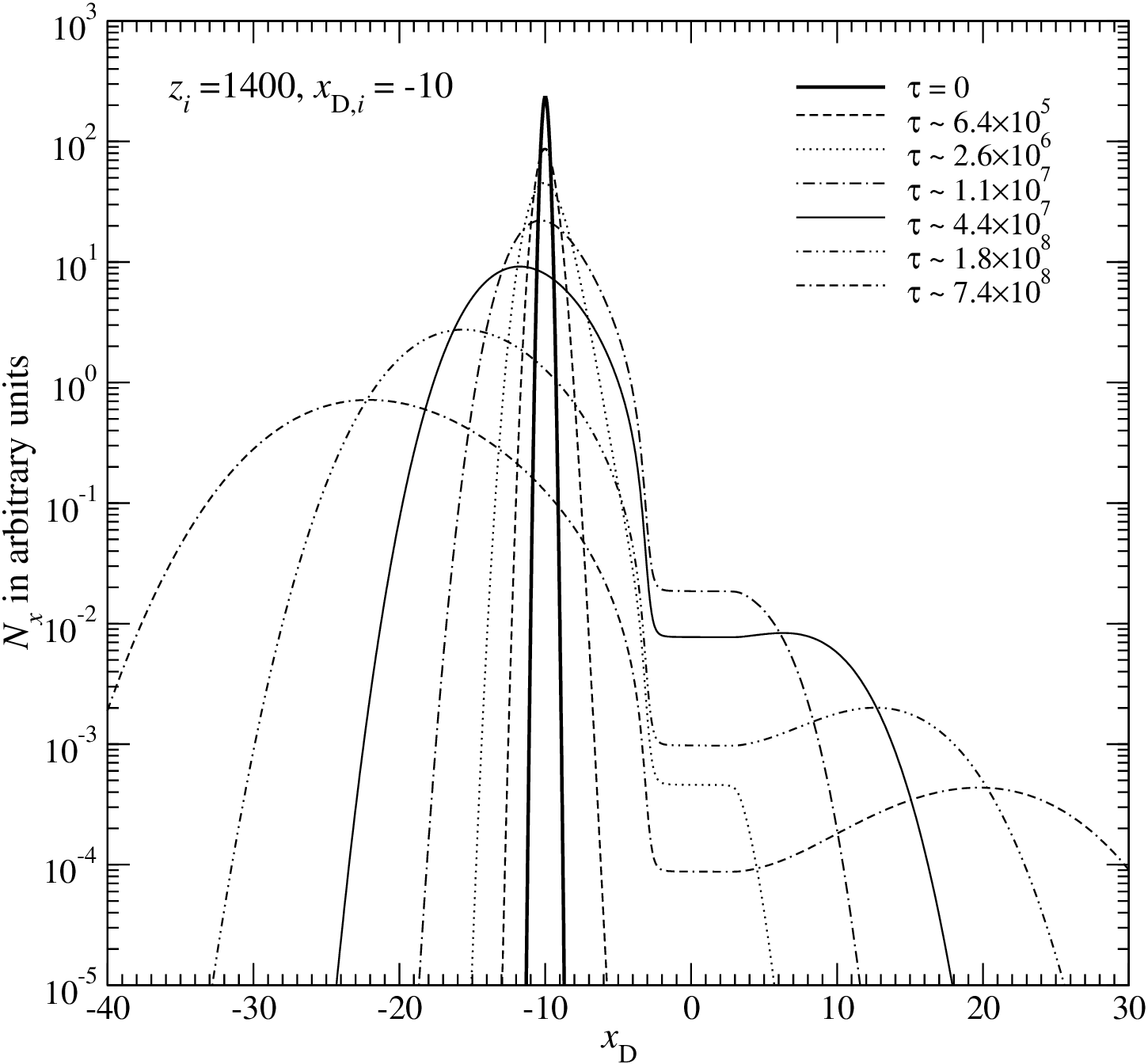}
\hspace{5mm}
\includegraphics[width=0.90\columnwidth]
{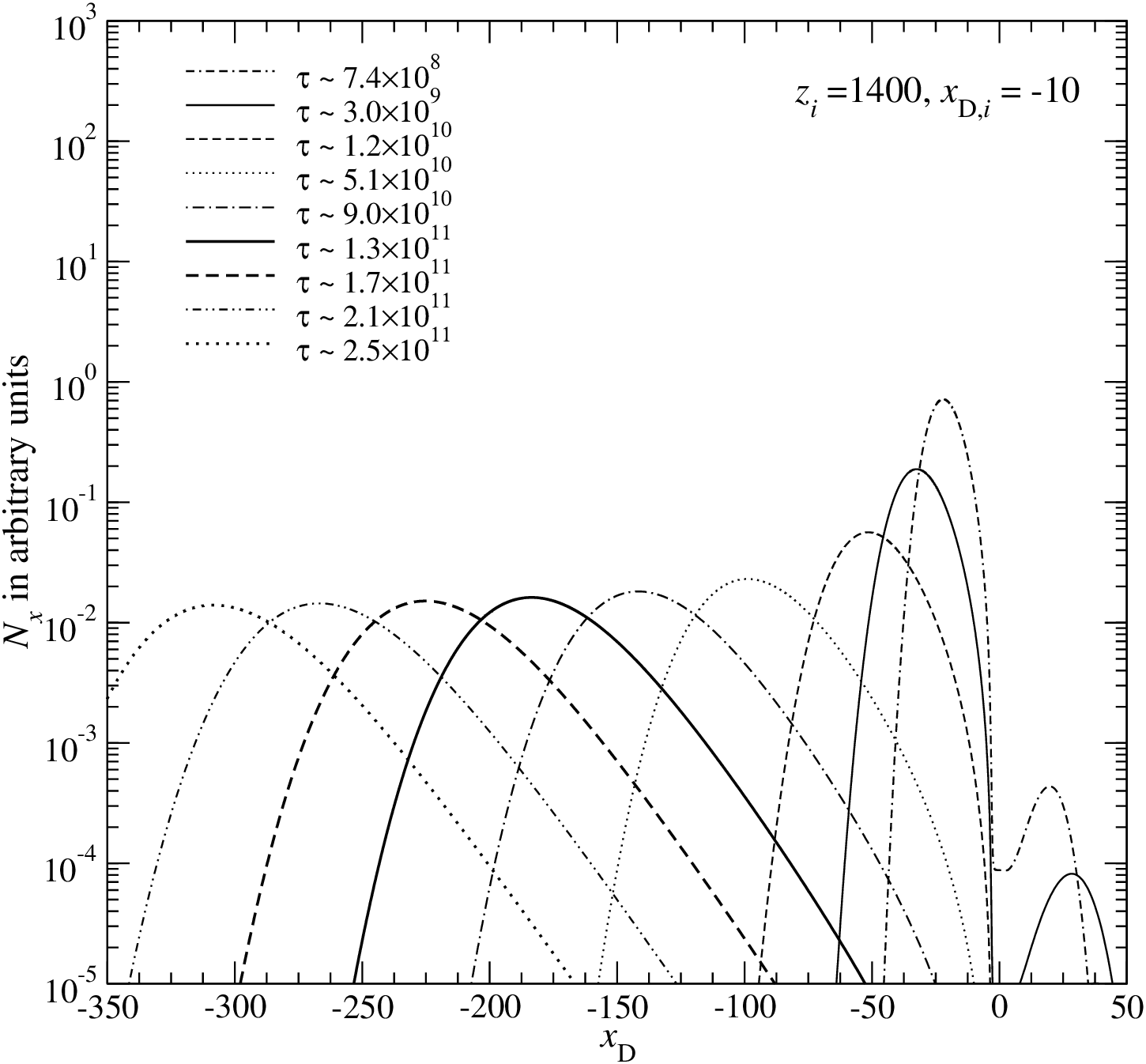}
\\
\includegraphics[width=0.90\columnwidth]
{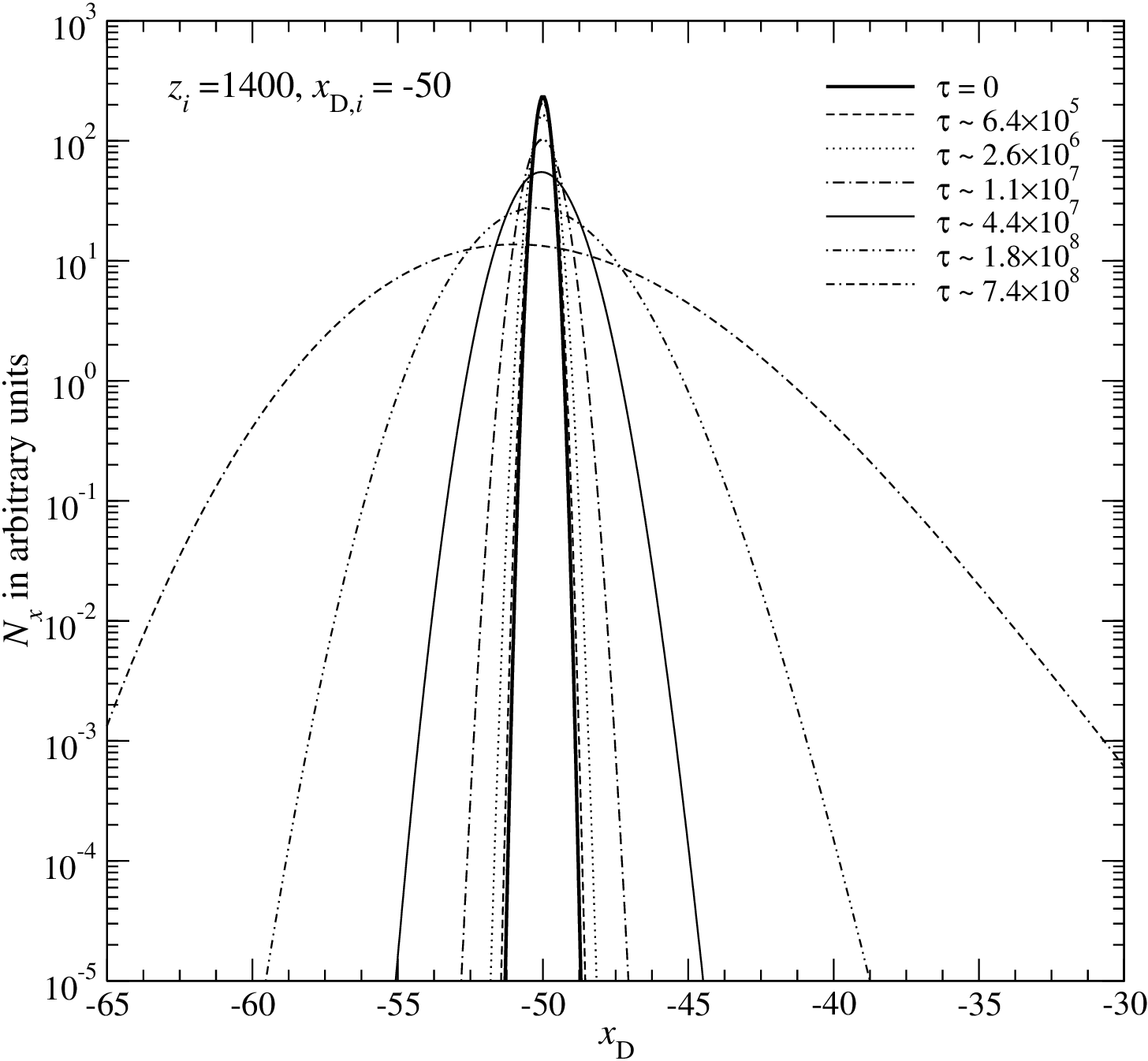}
\hspace{5mm}
\includegraphics[width=0.90\columnwidth]
{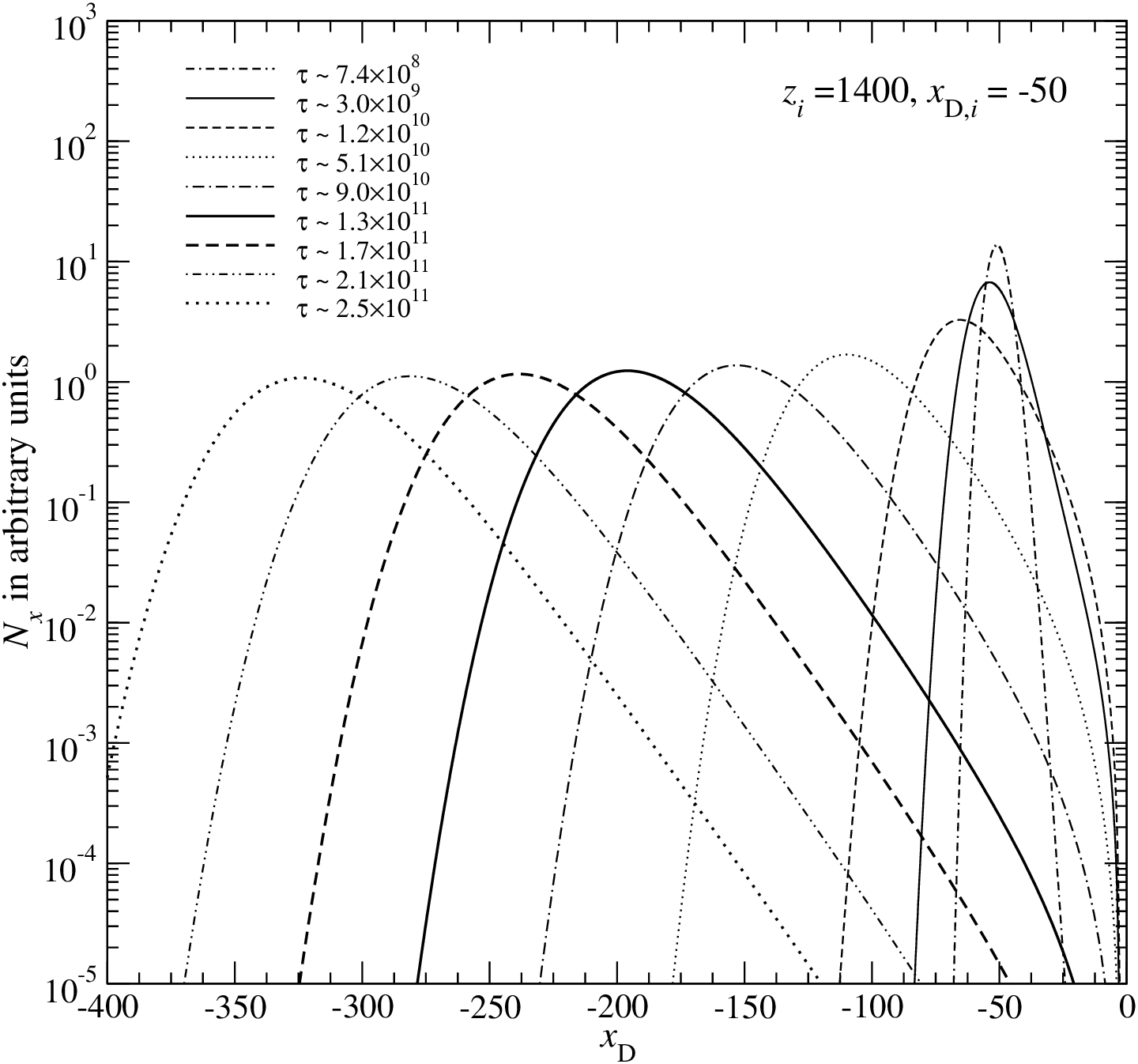}
\\
\includegraphics[width=0.90\columnwidth]
{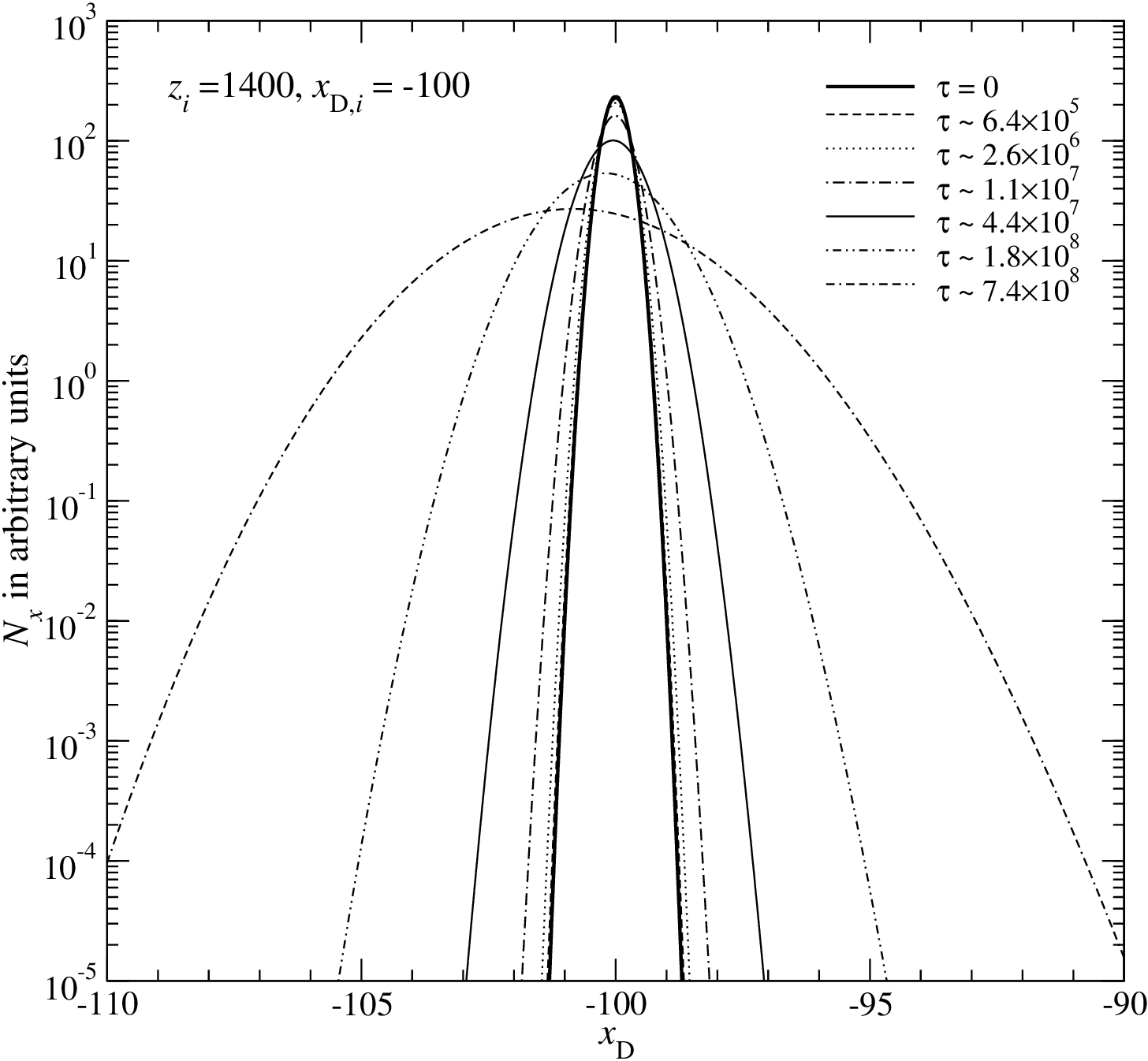}
\hspace{5mm}
\includegraphics[width=0.90\columnwidth]
{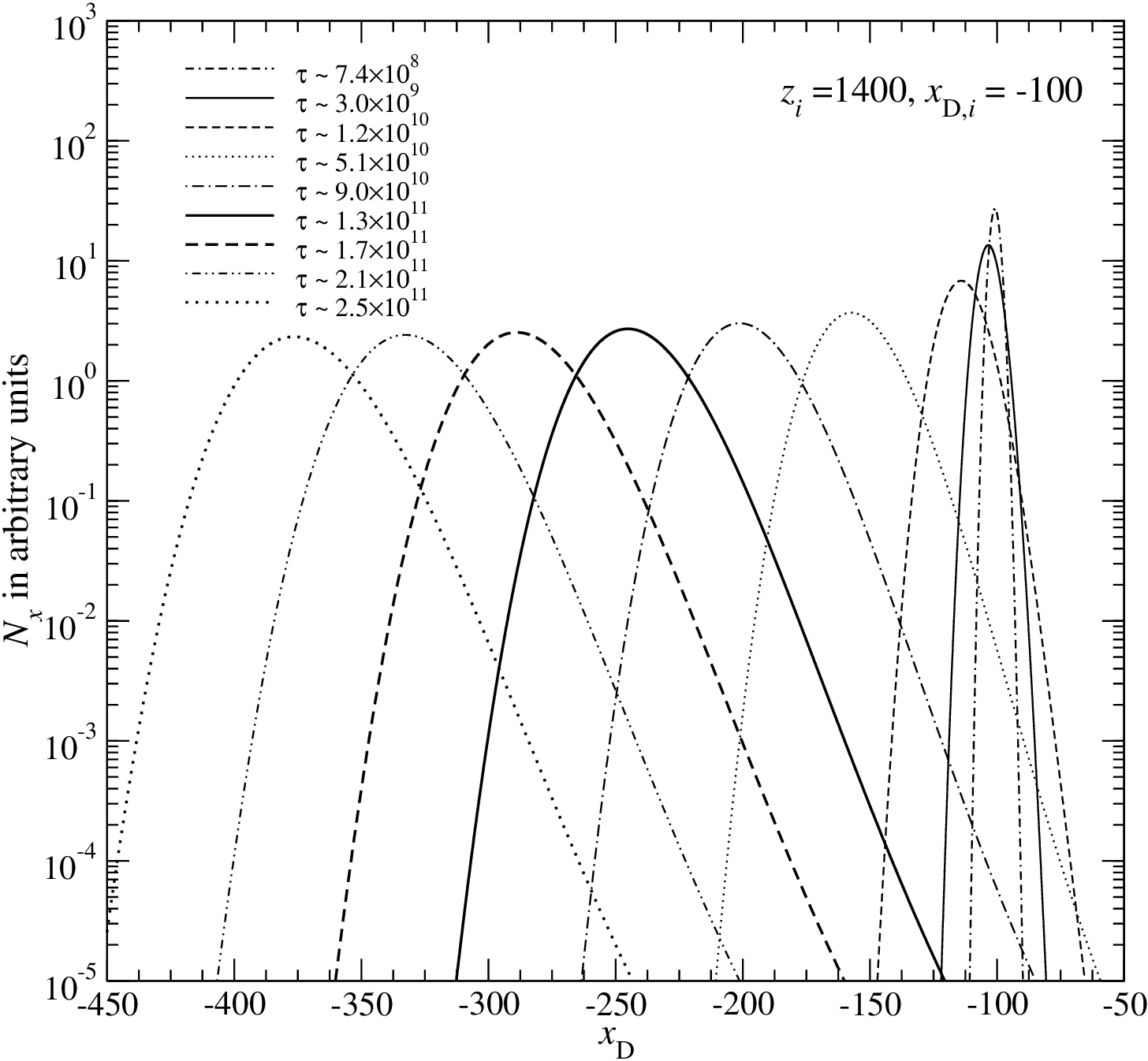}
\caption
{Time evolution of the photon distribution for single (narrow-line) injection
  on the red side of the Lyman-$\alpha$ resonance at different distance from
  the line center. The death probability for a 3-shell hydrogen atom was used
  and electron scattering has been neglected. We use the time-variable $\tau=\int c \,\sigma_{\rm r} N_{\rm H}\id t$.}
\label{fig:line_1400}
\end{figure*}

\begin{figure*}
\centering 
\includegraphics[width=0.90\columnwidth]
{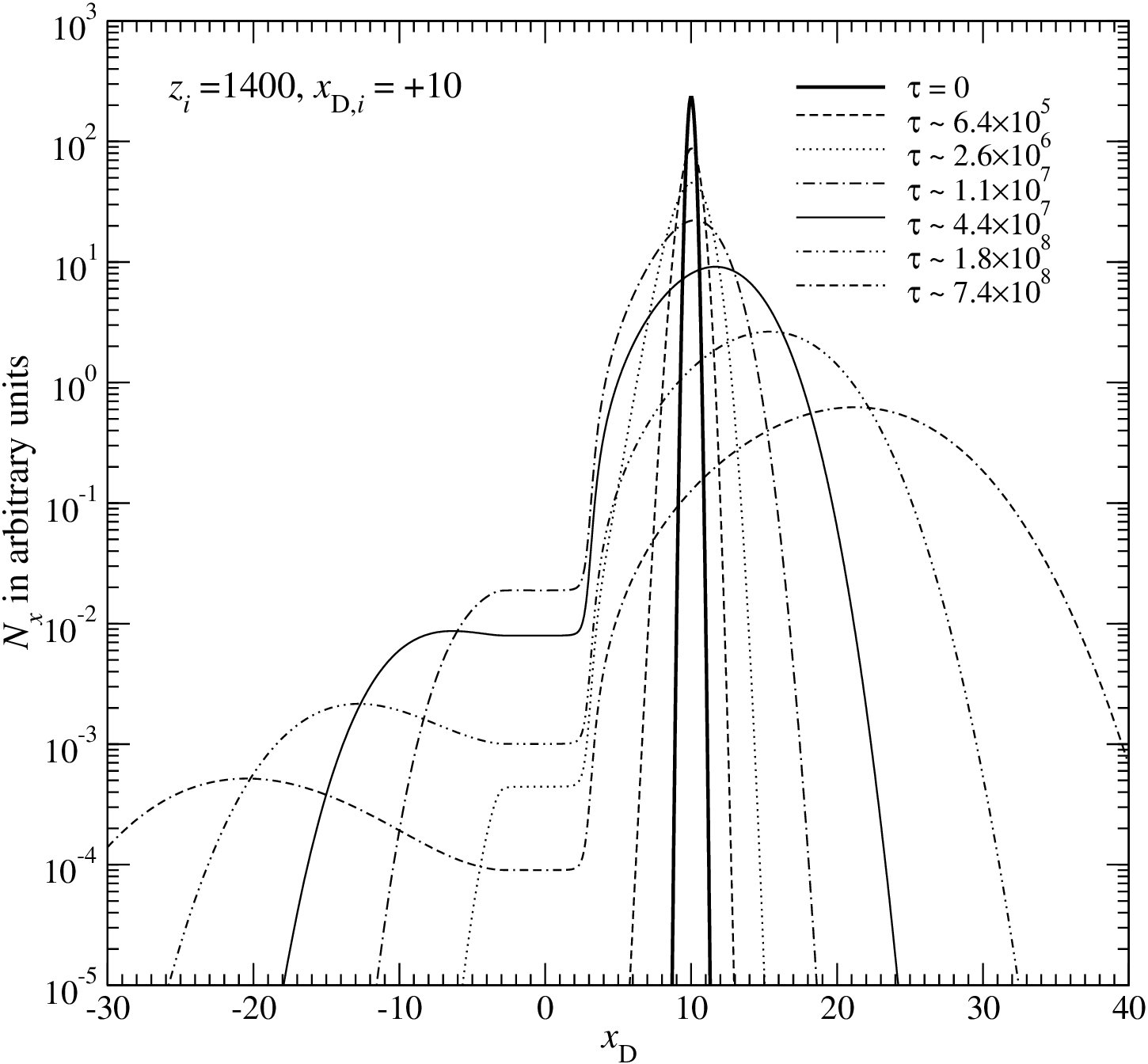}
\hspace{5mm}
\includegraphics[width=0.90\columnwidth]
{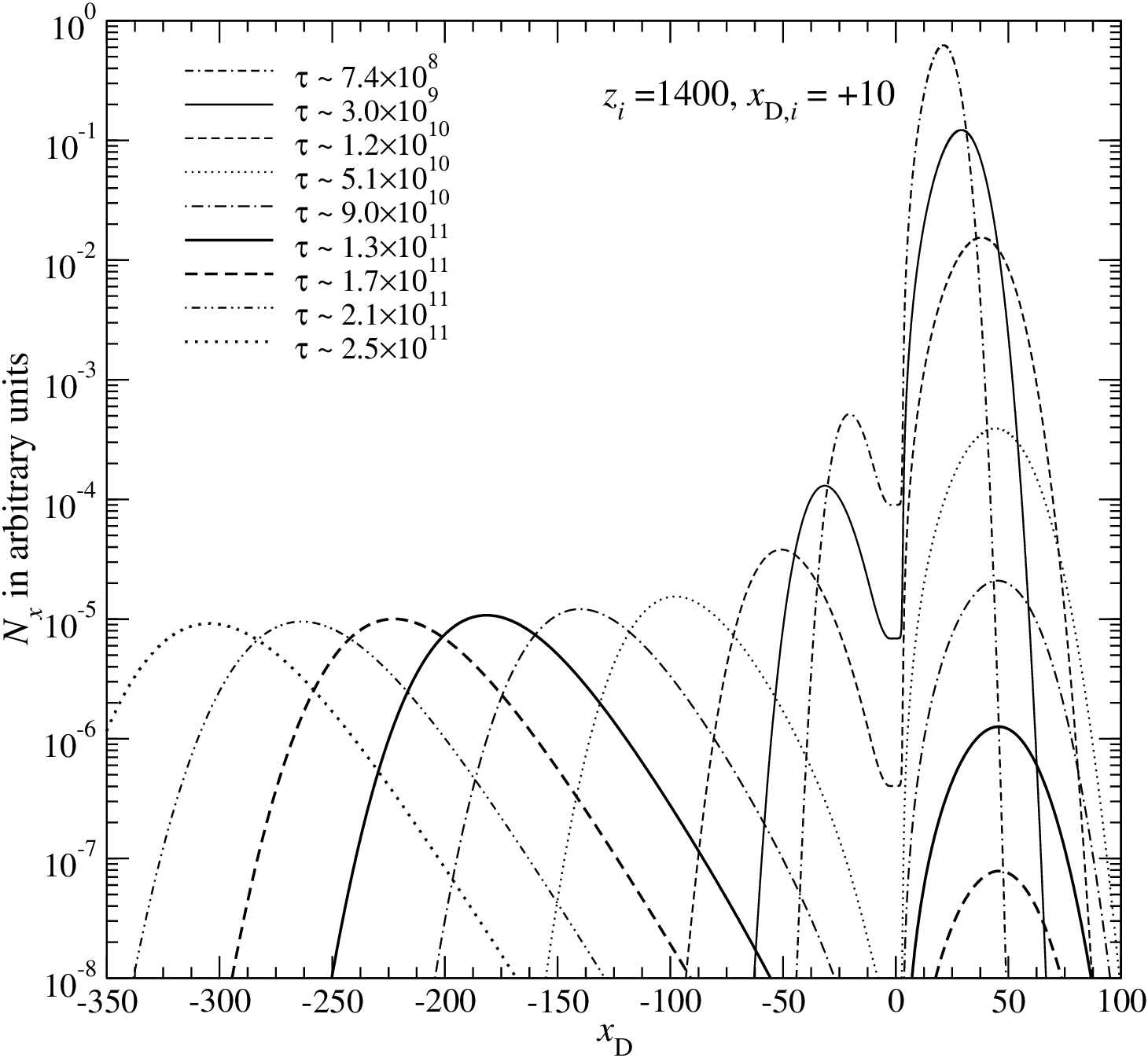}
\caption
{Time evolution of the photon distribution for single (narrow-line) injection
  at $\xD=+10$. The death probability for a 3-shell hydrogen atom was used and
  electron scattering has been neglected. We use the time-variable $\tau=\int c \,\sigma_{\rm r} N_{\rm H}\id t$.}
\label{fig:line_1400_xD_10}
\end{figure*}
\subsection{Time-dependent solutions}
\label{sec:narrow_line_inj}
\label{sec:narrow_line_inj_time}
In Fig. \ref{fig:line_1400_xD_0} we present the results for single injection of
photons at the Lyman-$\alpha$ line center. 
In practice we use a Gaussian initial photon distribution which is centered at
the injection frequency $x_{\rm D, i}$ and has a width $\sigma^2\sim \pot{5}{-2}$.
Furthermore, we re-normalized by a convenient factor such that induced effects
are negligible.
We started our computation at injection redshift $\zi=1400$, i.e. close to the time where the maximum of the CMB spectral distortion due to the Lyman-$\alpha$ transition appears
\citep{Jose2006}.
At this redshift roughly 20\% of all hydrogen atoms have already recombined
and the death probability for a 3-shell hydrogen atom\footnote{\change{The main contribution to the death of photons is due to the two-photon absorption to the 3d-state. Including more shells the death probability changes by less that 10\% during hydrogen recombination. \citep{Chluba2009}.}} is $p_{\rm d}\sim
\pot{5.6}{-4}$ \citep[see Fig. 1 in][]{Chluba2008b}.

From Fig. \ref{fig:line_1400_xD_0} one can see that after a short time the
initial photon distribution has broadened significantly, bringing
photons to the {\it wings} of the Lyman-$\alpha$ transition. After
$\tau=\int c\,\sigma_{\rm r}\,N_{\rm H} \id t\sim 10^4$ the death of photons
in the line center becomes important, owing to the fact that $p_{\rm d}\sim 10^{-4}$ is so
small.
The solution remains very symmetric until $\tau\sim \text{few}\times 10^{10}$
and only then redshifting due to the expansion of the Universe starts to
become important (as we will see line-recoil only affects the photon
distribution at the level of few percent in addition).
%
When the bulk of photons reaches a distance $\xD\sim -100$ still a sizable
amount of them remains on the blue side of the Lyman-$\alpha$ line, and only
when the maximum of the photon distribution reaches $\xD\sim -150$ the
evolution starts to become dominated by redshifting and absorption only, with
very small changes because of frequency redistribution.

In Fig. \ref{fig:line_1400} we present the results for single injection of
photons at different distances to the line center. Again photons were injected
at $\zi=1400$.
Focusing on the case $x_{\rm D, i}=-10$, one can again observe the fast
broadening of the initial photon distribution. However, now the characteristic
time for line scattering has increased by a factor of $\sim \pot{2.3}{5}$
because frequency redistribution already takes place in the wings of the Voigt-profile.
It is important to note that due to line scattering photons strongly diffuse
back into the line center and thereby increase the possibility of being
absorbed. Also one can see that due to diffusion some photons even reach far
into the blue side of the Lyman-$\alpha$ resonance.
Again only after the bulk of photons has reached a distance of $\xD\sim -150$
redshifting and absorption play the most important role for the evolution of
the photon distribution.

Looking at the other two cases, it becomes clear that for injection at $x_{\rm
  D, i}=-50$ still a few photons do diffuse back to the line center, whereas
  for $x_{\rm D, i}=-100$, practically all photons remain below $\xD\sim-50$
  at all times.
Comparing the maxima of the final photon distribution (at $\tau\sim
\pot{2.5}{11}$) for all the discussed cases shows that as expected the
efficiency of absorption decreases when increasing $x_{\rm D, i}$.

It is also interesting to look at cases when injecting photons on the {\it
  blue} side of the Lyman-resonance. In this case all photons have to pass at
least once through the resonance before they can escape and one expects that
many photons die during this passage.
In Fig. \ref{fig:line_1400_xD_10} we show the results for single injection at
$\xD=+10$. At the beginning the evolution of the spectrum looks very similar
(except for mirror-inversion) to the case of injection at $\xD=-10$. However,
at late times one can see that the amount of photons reaching the red side of
the Lyman-$\alpha$ resonance is significantly smaller. Indeed this amount is
comparable to the case of injection directly at the center.
%

\subsection{Escape probability for single narrow line injection}
\label{sec:Pesc_def_num}
Given an initial photon distribution one can compute the total number
of photons that {\it survive} the evolution over a period of time for the given transfer problem.
Here we assume that only at time $t=0$ fresh photons are appearing.  
Comparing the total number of photons at the final stage with the initial
number then yields the {\it numerical escape} or {\it survival probability}
for the given diffusion problem
\beal
\label{app:P_esc_def}
P_{\rm esc}(z_{\rm i}, z_{\rm f})=
\frac{N_\gamma(z_{\rm f})}{N_\gamma(z_{\rm i})}\left[\frac{1+z_{\rm i}}{1+z_{\rm f}}\right]^3
\equiv
\frac{\int\tilde{N}_{x}(z_{\rm f})\id x}{\int\tilde{N}_{x}(z_{\rm i})\id x},
\end{align}
where $N_\gamma(z)=\int N_\nu(z)\id\nu$ is the number density of photons at
redshift $z$. The factors $(1+z)^3$ account for the changes in the scale
factor of the Universe between the initial and final redshift.

Due to the expansion of the Universe photons redshift towards lower
frequencies. Neglecting any redistribution process, with time this will
increase the distance of the initial photon distribution to the line center
and thereby decrease the probability of real line absorption.
Assuming that the initial photon distribution is given by a $\delta$-function then with
Eq.~\eqref{app:P_esc_def} one obtains 
\beal
\label{app:P_esc_delta}
P^{\delta, \rm abs}_{\rm esc}=e^{-\tau_{\rm abs}(\nu, z_{\rm i}, z)}
\end{align}
for this case. Here $\tau_{\rm abs}$ is the absorption optical depth between the initial redshift $z_{\rm i}$ and $z$. 

We now want to compare the differential escape probability Eq.~\eqref{app:P_esc_delta} with the numerical results obtained when also including the redistribution of photons over frequency. 
The results of the the previous Section suggest the following: 
\begin{itemize}
\item[(i)] For photons injected close to the line center the diffusion due to
resonance scattering {\it helps} to bring photons towards the wings. In
comparison to the case with no scattering this should increase the escape
probability.
  
\item[(ii)] At intermediate distances on the {\it red} side of the line center
  ($\xD\sim -50$ to $-100$ Doppler width) line diffusion brings some photons back to the Doppler core
  and thereby should decrease the escape probability in comparison to the case
  without line scattering. 
    
\item[(iii)] Far in the {\it red} wing of the line ($\xD\lesssim -100$) the
  escape fraction will depend mainly on the death probability and the expansion rate of
  the Universe. In this regime line scattering does lead to some line
  broadening, but should not affect the escape probability
  significantly anymore.
  
\item[(iv)] The escape probability for injections on the {\it blue} side of
  the resonance becomes nearly independent of the initial distance to the line
  center and should be comparable to the one inside the Doppler core.

\end{itemize}

It is easy to check these statements numerically. For this we performed a
sequence of computations injecting photons at different distances from the
line center and following their evolution until the initial maximum of the
photon distribution has reached $x_{\rm D, t}= x_{\rm D, i}-x_{\rm D, s}$.
We then computed the escape or survival probability as defined by
Eq.~\eqref{app:P_esc_def} for the given diffusion problem as a function of the
injection frequency, $x_{\rm D, i}$, injection redshift, $z_{\rm i}$, and
termination redshift, $z_{\rm t}$, which directly depends\footnote{For
simplicity we used $z_{\rm t}=z_{\rm i}[1+x_{\rm D, t}\Delta\nu_{\rm D}(z_{\rm
i})/\nu_{21}]$.} on the value of $x_{\rm D, s}$.

Since the absorption cross section in wings of the line scales like $\propto
1/\xD$, even
beyond $\xD\sim -10^3$ still percent-level absorption can occur, which should
be taken into account when computing the total escape probability until
redshift $z_{\rm t}=0$. However, the effect of resonance scattering becomes
negligible at this distance from the line center (see below) and the time
evolution in principle can be described fully analytically.  
For simplicity we neglected this additional complication
and typically chose $x_{\rm D, s}\sim 10^4$, which ensured that the remaining
absorption will only lead to modifications of $\Delta P/P\lesssim 10^{-3}$ to
the obtained escape probability. Up to this level of accuracy, the obtained curves presented in this Section can
be considered as the frequency-dependent total escape probability until
$z_{\rm t}=0$.

\begin{figure}
\centering 
\includegraphics[width=0.9\columnwidth]
{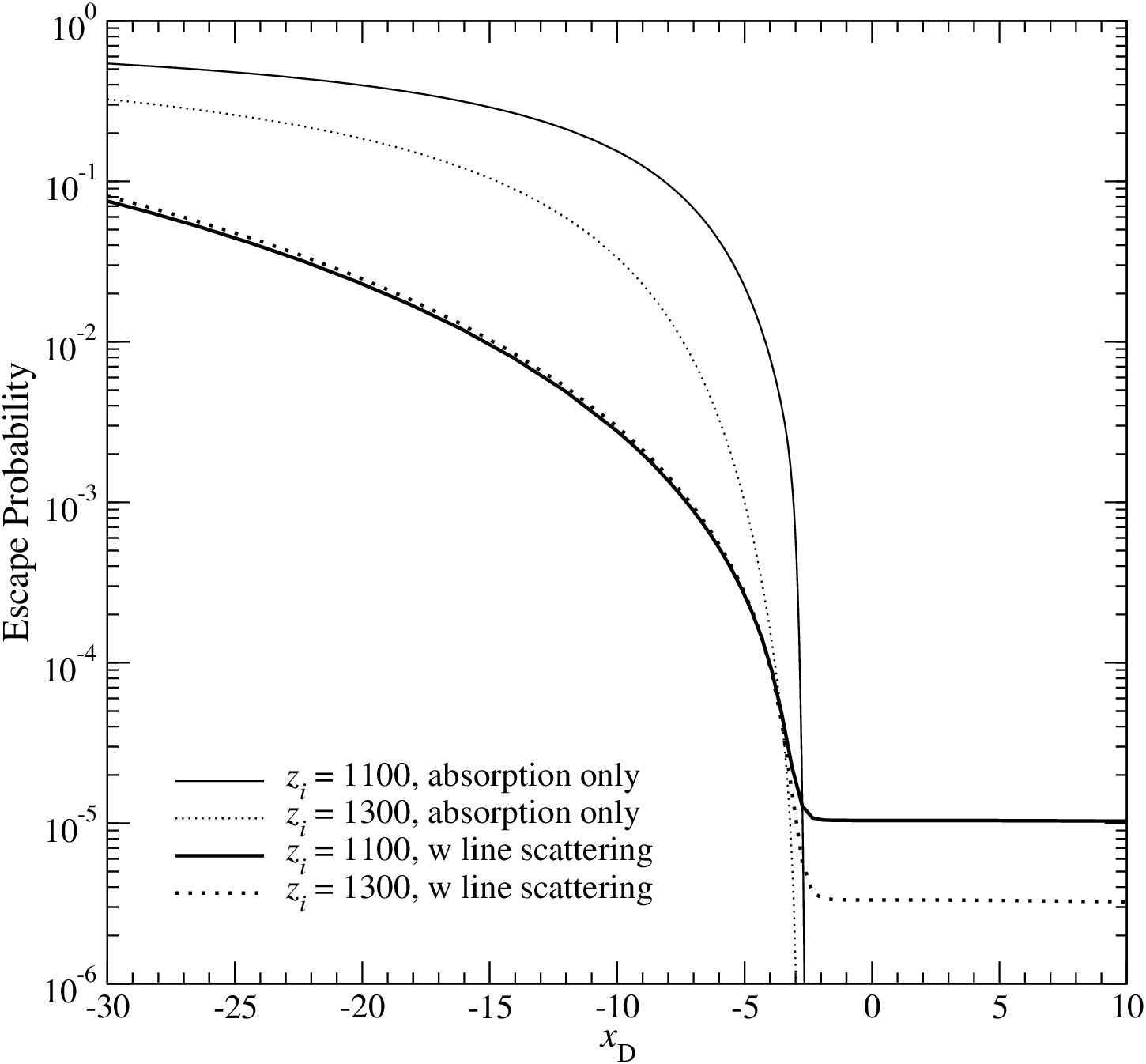}
\\
\includegraphics[width=0.9\columnwidth]
{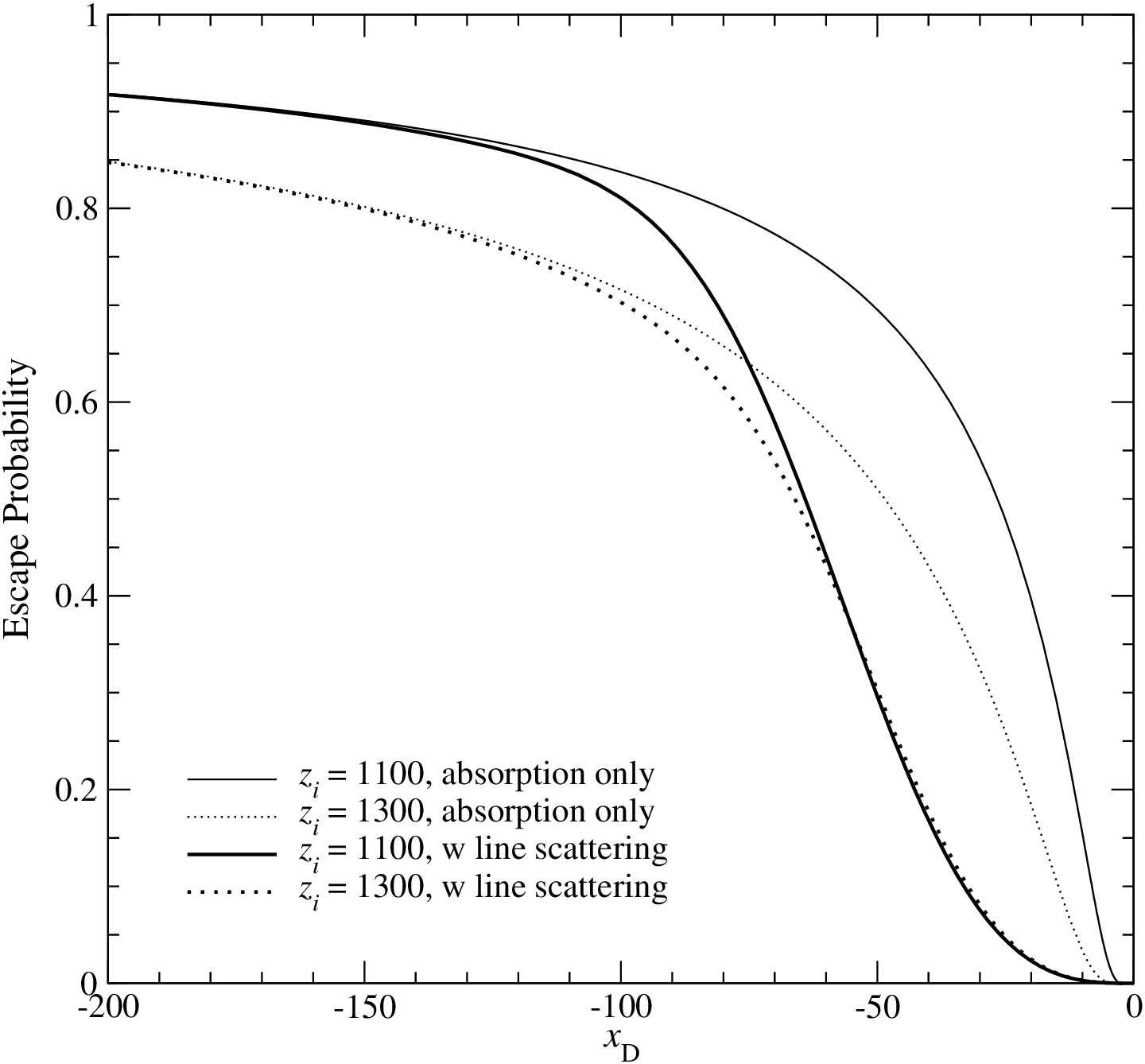}
\caption
{Escape probability, $P_{\rm esc}(\xD, z_{\rm i}, z_{\rm t})$, for single
  (narrow line) injection at different distances from the line center and initial
  redshifts $z_{\rm i}=1100$ and $1300$. The death probability for a 3-shell hydrogen atom
  was used and electron scattering has been neglected. For the given curves we
  set $x_{\rm D, s}= 10^4$, such that $P_{\rm esc}(\xD, z_{\rm i}, z_{\rm
    t})\approx P_{\rm esc}(\xD, z_{\rm i}, 0)$. For comparison also the
  analytic result, $P^{\delta\rm, abs}_{\rm esc}$, for $\delta$-function
  injection including only pure absorption without line scattering is shown.}
\label{fig:p_esc_line}
\end{figure}
In Fig. \ref{fig:p_esc_line} we present some results for computations of the
frequency-dependent escape probability, $P_{\rm esc}(\xD, z_{\rm i}, z_{\rm
  t})$, for injection redshifts $z_{\rm i}=1100$ and $1300$. For comparison we also give the
corresponding escape probabilities, $P^{\delta\rm, abs}_{\rm
  esc}=e^{-\tau_{\rm abs}}$, Eq.~\eqref{app:P_esc_delta}, for
$\delta$-function injection when neglecting line scattering.
At large distance ($\xD\lesssim -150$) from the line center $P_{\rm esc}$ practically coincides
with $P^{\delta\rm, abs}_{\rm esc}$ in all presented cases. As mentioned above
this behavior is expected since line scattering should not strongly affect the
evolution of the line anymore.
At intermediate distances from the line center the inclusion of line
scattering indeed decreases the escape probability in comparison to the cases
without scattering. 
Looking in detail at the dependence of $P_{\rm esc}$ close to the center of
the line shows that the presumptions (i) and (iv) also hold.  Our computations
clearly show that there is a non-vanishing escape probability for photons from
the blue side of the line, which in the case of pure absorption is practically
zero\footnote{There is a small difference close to the line center due to the fact
  that we used $\delta$-function injection for the computation of
  $P^{\delta\rm, abs}_{\rm esc}$ instead of the Gaussian that was used in the numerical computation. However, this will only make the transition
  $P^{\delta\rm, abs}_{\rm esc}\rightarrow 0$ less steep, but otherwise will
  not change the main conclusion.}. This probability is nearly constant
extending even into the core of the line and down to $\xD\sim -2$.

\begin{figure}
\centering 
\includegraphics[width=0.9\columnwidth]
{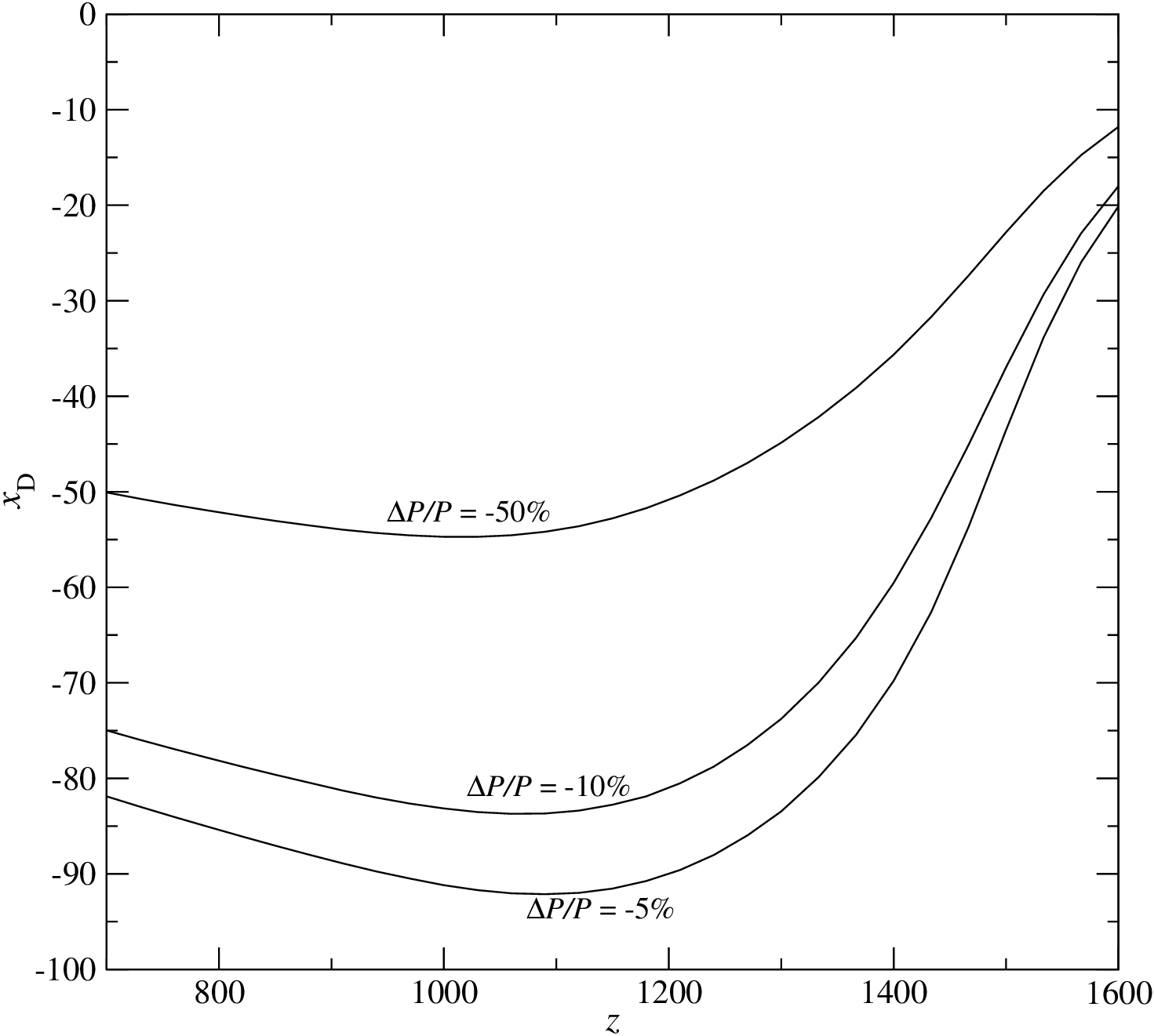}
\caption
{Frequency $\xD<0$ at which the modification due to line scattering becomes
  $\epsilon$ percent. Here $\Delta P/P\equiv [P^{\rm sc}_{\rm
  esc}-P^{\delta\rm, abs}_{\rm esc}]/P^{\delta\rm, abs}_{\rm esc}$, where for
  $P^{\rm sc}_{\rm esc}$ line scattering was taken into account.
The death probability for a 3-shell hydrogen atom was used and we set $x_{\rm
D, s}=10^4$ for all curves.}
\label{fig:xDc.scatt}
\end{figure}
In order to understand up to which distance to the line center the effect of
resonance scattering is important we compared the results for the escape
probability including line scattering with the analytic no-scattering
solution, asking the question at which distance in the red wing the
modification due to line scattering becomes $\epsilon$ percent.
In Fig.~\ref{fig:xDc.scatt} we summarize the results of such comparison.
It is clear that at all redshifts of interest line-scattering is only
important for $\xD\gtrsim -\text{few}\times10^2$, but at the percent-level in
principle may be neglected below this frequency.
We made use of this result already in some earlier works \citep{Chluba2008a}.

\subsubsection{Role of atomic-recoil}
\label{sec:pesc_linerecoil}
Every resonance scattering due to atomic-recoil leads to a small shift of the
photon energy towards lower frequencies.
The strength of the recoil due to the frequency-dependence of the scattering
cross section is a strong function of photon energy, peaking close to the
Lyman-$\alpha$ line center, and dropping rather strongly in the damping wings.
This is in stark contrast to electron-recoil, for which the scattering cross
section is practically independent of frequency.

To understand the importance of the atomic recoil effect for the differential escape probability we therefore performed several computations of the frequency-dependent escape
probability for injection of photons at different distances from the line
center explicitly neglecting the effect of atomic recoil.
%
\begin{figure}
\centering 
\includegraphics[width=0.9\columnwidth]
{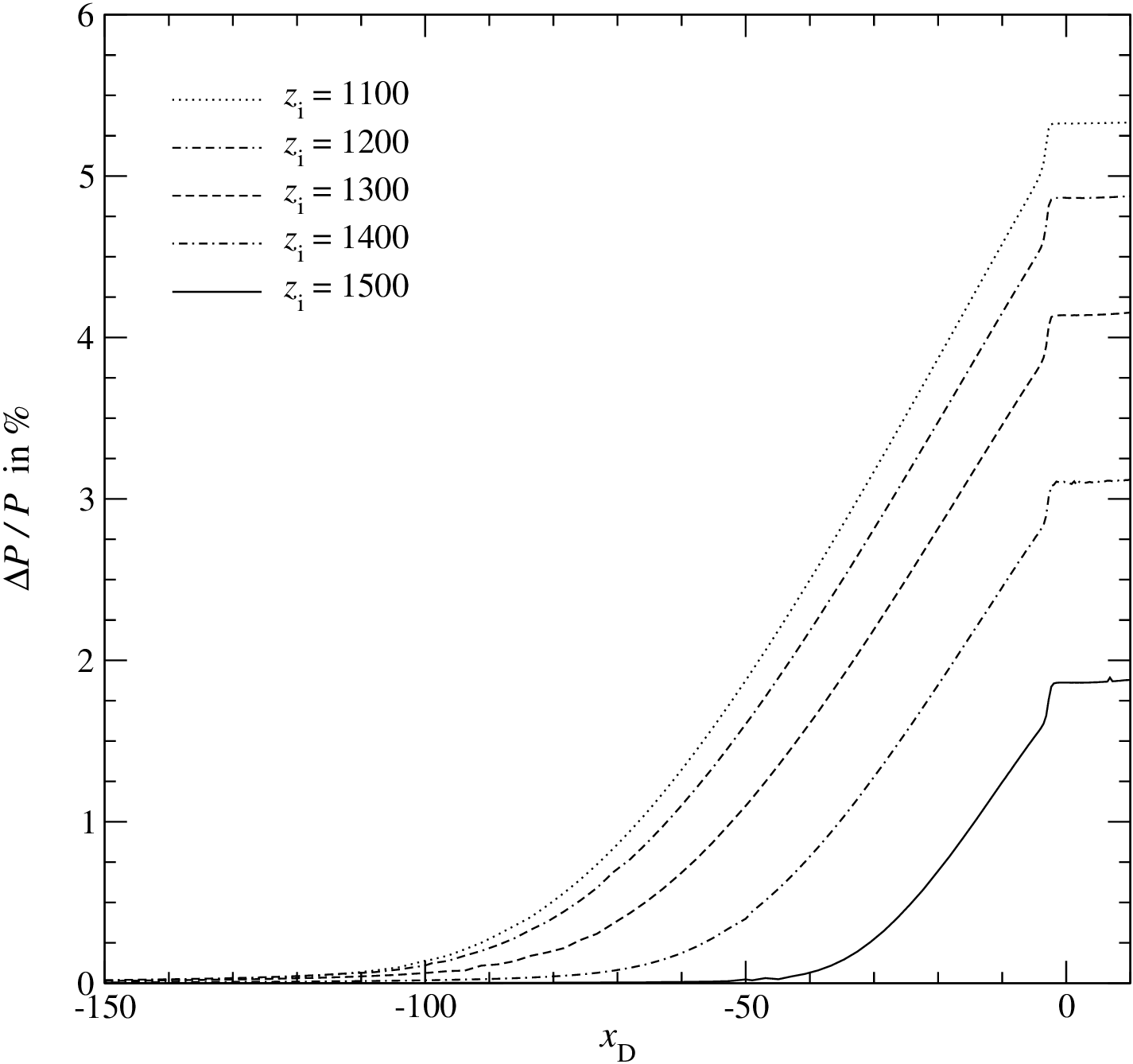}
\caption
{Relative difference in the escape probability for single (narrow line)
  injection at different distances from the line center when including the
  effect of atomic recoil. Here $\Delta P/P\equiv [P_{\rm esc}-P^{\rm
  no-rec}_{\rm esc}]/P^{\rm no-rec}_{\rm esc}$, where for $P^{\rm no-rec}_{\rm
  esc}$ the term due to atomic recoil was neglected.
The death probability for a 3-shell hydrogen atom was used and we set $x_{\rm
D, s}=10^4$.}
\label{fig:Dpesc.L.var_z.recoil}
\end{figure}
In Fig.~\ref{fig:Dpesc.L.var_z.recoil} we present the correction to the escape
probability which is only due to the atomic-recoil term.
As expected, atomic recoil helps photons to escape in the whole range of
frequencies. However, due to the decrease in the scattering cross-section the
corresponding correction becomes very small at distances below $\xD\sim -100$
to $-150$.
Also the amplitude of the effect increases towards lower redshifts, simply
because more hydrogen atoms have become neutral.
The largest correction is coming from the line center and in practically constant over the whole Doppler core and the blue side of the resonance.
We will see below that the total correction to the Lyman $\alpha$ escape probability is very similar to the value obtained for injections close to the line center (see Sect.~\ref{sec:Lya_Pesc}).

\subsubsection{Role of electron scattering}
\label{sec:pesc_e_scatt}
\begin{figure}
\centering 
\includegraphics[width=0.9\columnwidth]
{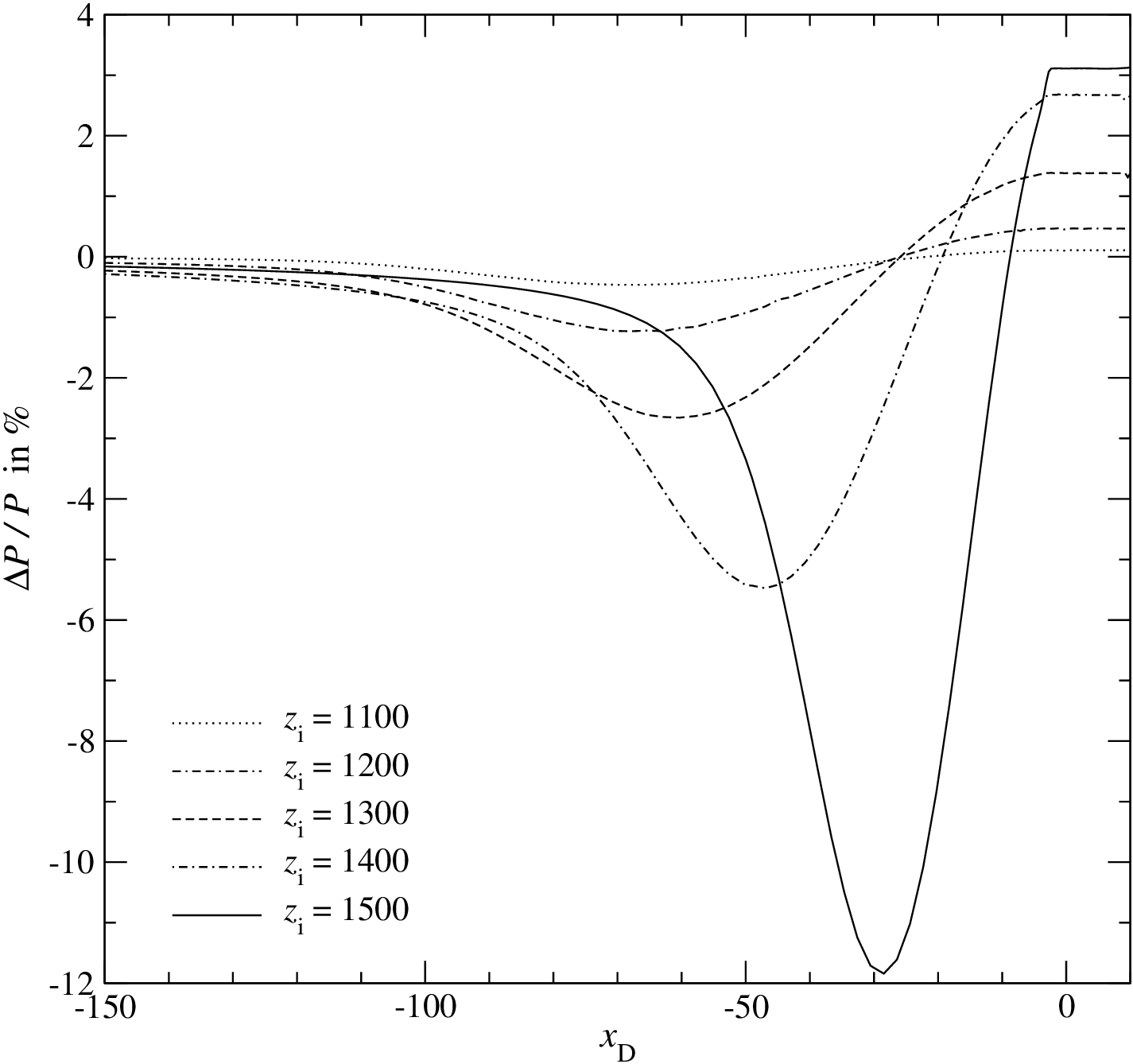}
\caption
{Relative difference in the escape probability for single (narrow line)
  injection at different distances from the line center when including
  electron scattering. Here $\Delta P/P\equiv [P^{\rm e}_{\rm esc}-P_{\rm
    esc}]/P_{\rm esc}$, where for $P^{\rm e}_{\rm esc}$ electron scattering
  was taken into account.
The death probability for a 3-shell hydrogen atom was used and we set $x_{\rm
D, s}=10^4$ for all curves.}
\label{fig:Dpesc.L.var_z.e_scatt}
\end{figure}
In Fig. \ref{fig:Dpesc.L.var_z.e_scatt} we show the relative difference in the
escape probability for single (narrow line) injection at different distances
from the line center when including electron scattering.
As expected electron scattering has a similar effect as resonance scattering,
helping photons to escape more efficiently from the line center, but bringing
some photons from the wings back into the Doppler-core, diminishing the
probability of their survival.
At higher injection redshift the differences become larger, due to the
increase of the number of free electrons as compared to the number of neutral
hydrogen atoms. At $\zi\lesssim 1200$ the relative difference becomes smaller
than $\sim 1.2\%$ in the whole considered range of injection frequencies.
Close to the maximum of the visibility function $\zi\sim 1100$ one does not
expect a large correction due to electron scattering. 
In addition it is clear that the increase in the escape in the Doppler-core
should be partially canceled by the decrease in the red wing.
As we will see in Sect.~\ref{sec:Lya_Pesc} the net effect of electron
scattering on the Lyman $\alpha$ escape probability during hydrogen recombination is always $\lesssim
1\%$ at $z\lesssim 1600$.

\section{Changes in the Lyman $\alpha$ escape probability during hydrogen recombination}
\label{sec:Lya_Pesc}
In this Section we now present the results for the changes in the Lyman $\alpha$ escape probability during hydrogen recombination. 
Our approach here is very similar to the one used in our earlier, semi-analytical works \citep{Chluba2008b, Chluba2009}.
Given the solution for the populations of the different hydrogen levels we numerically solve the transfer equation for the Lyman $\alpha$ problem obtaining the spectral distortion in the vicinity of the Lyman $\alpha$ resonance at different redshifts. 
From this we can compute the effective escape probability by convolving this distortion with the corresponding Lyman $\alpha$ absorption profile.
\change{We also followed a very similar approach in our previous computations of the radiative transfer problem during helium recombination, where some of the results obtained in that case were already used in \citet{Jose2008}.}

We will start by discussing the results in the standard '$1+1$' formulation (Sect.~\ref{sec:phi_V}). 
We then include the effect due to the {\it thermodynamic correction factor} $f_\nu$ (Sect.~\ref{sec:fnu}), which was introduced earlier \citep{Chluba2008b, Chluba2009} using the detailed balance argument.
Finally we shall also include the corrections to the 3d-1s and 3s-1s two-photon emission profile (Sect.~\ref{sec:phi_2g}).

\begin{figure}
\centering
\includegraphics[width=0.9\columnwidth]{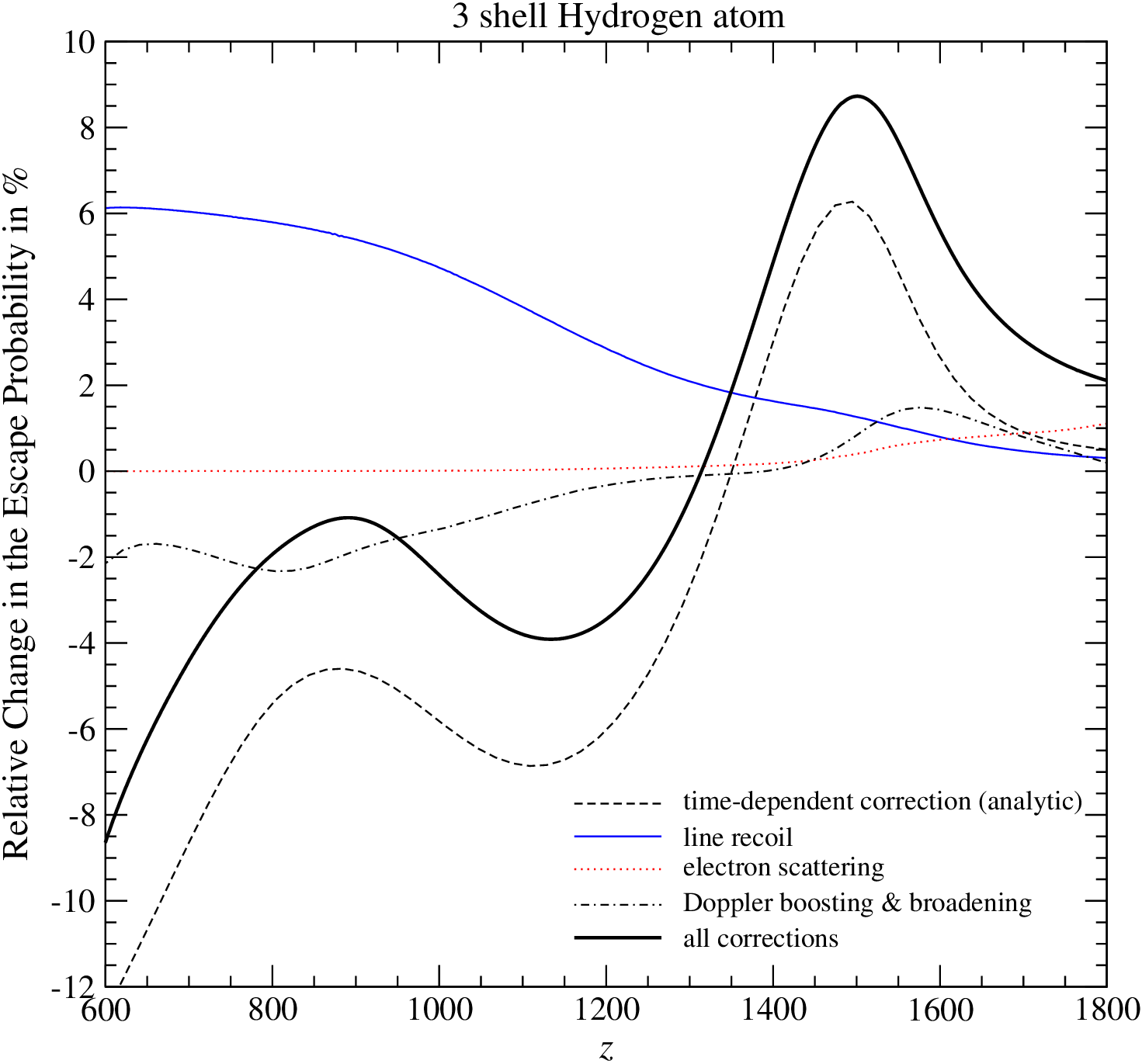}
\caption
{Changes in the Lyman $\alpha$ escape probability for the standard '$1+1$' photon formulation. The dashed line shows the result obtained in the no scattering approximation \citep{Chluba2009}.}
\label{fig:DP_phi_V}
\end{figure}
\subsection{Results in the standard '$1+1$' photon formulation}
\label{sec:phi_V}
In Figure~\ref{fig:DP_phi_V} we present the results for the escape probability using the standard '$1+1$' photon formulation. In this case the emission and absorption profile are given by the normal Voigt profile.
\change{We also included the full time-dependence of the problem in the computations of the line emission rate and the absorption optical depth. In the no scattering approximation \citep{Chluba2008b} this leads to the dashed curve shown in Fig.~\ref{fig:DP_phi_V}.}

As mentioned earlier \citep{Chluba2008b} the standard '$1+1$' photon formulation has several discrepancies, i.e. leading to an unphysical {\it self-feedback} of Lyman $\alpha$ photons at low redshifts ($z\lesssim 800-900$). 
Nevertheless, one can study the influence of the redistribution of photons by resonance and electron scattering even in this approach and as we will see one obtains very similar results for the effect of resonance scattering in comparison with the more complete formulation using the two-photon picture (Sect.~\ref{sec:phi_2g}).

In Figure~\ref{fig:DP_phi_V} we show the separate correction due to atomic recoil (thin solid line). We obtained this curve by taking the difference of the escape probabilities for the case with all corrections due to line and electron scattering included and the one in which line recoil was switched off.
The importance of recoil increases towards lower redshifts reaching the level of $\Delta P/P \sim 6\%$ at $z\sim 600$. 
%
%
If we look at the results presented in Fig.~\ref{fig:Dpesc.L.var_z.recoil} for the case of single narrow line injection, we can even see that the total recoil correction seen in Fig.~\ref{fig:DP_phi_V} is very close to the value obtained for line center injection.
This is expected, since the largest contribution to the total value of the escape probability always comes from the Doppler core.
%

\begin{figure}
\centering
\includegraphics[width=0.9\columnwidth]{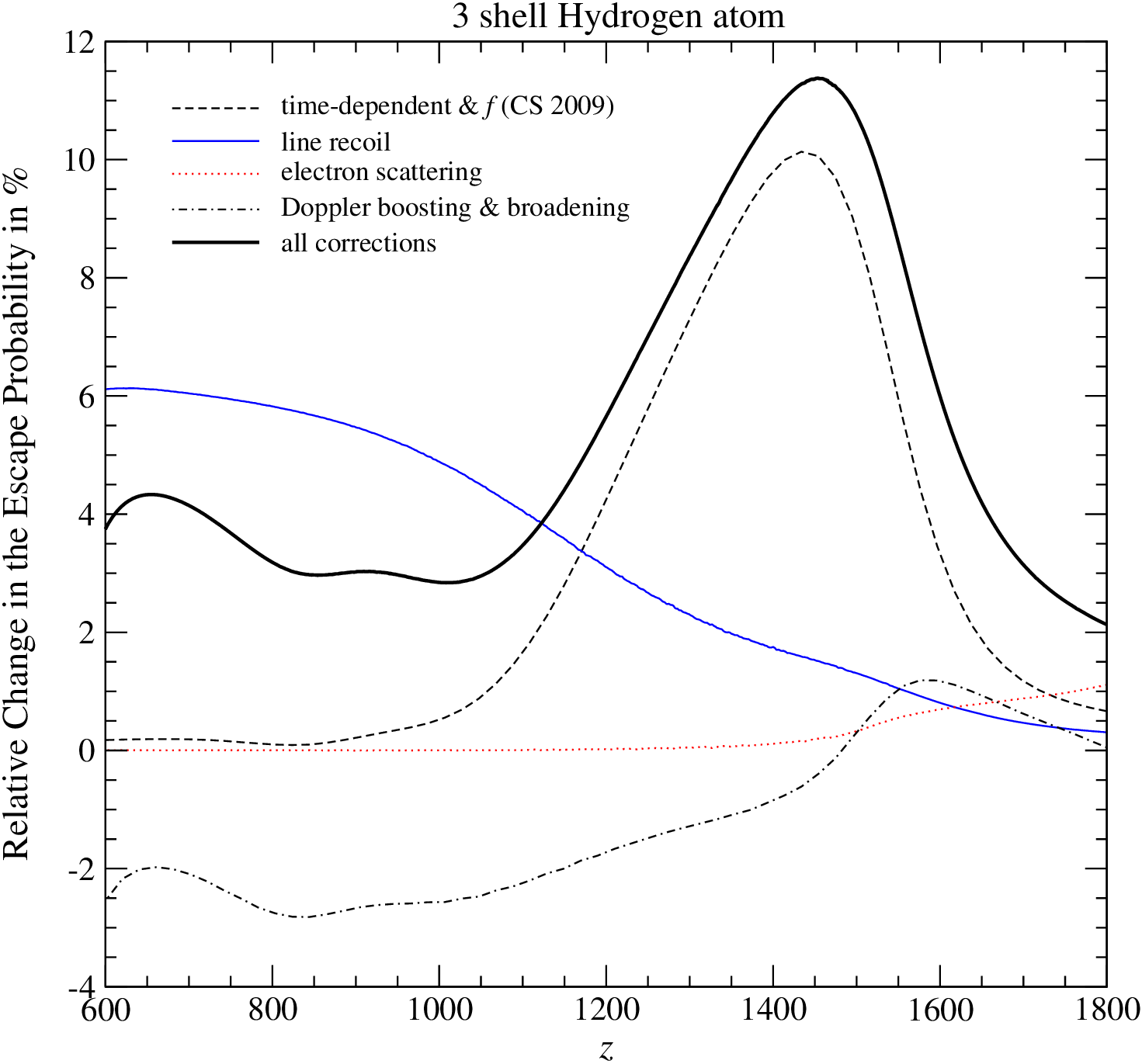}
\caption
{Changes in the Lyman $\alpha$ escape probability due to the thermodynamic correction factor. The dashed line shows the result obtained in the no scattering approximation \citep{Chluba2009}.}
\label{fig:DP_fnu}
\end{figure}
We can also see that the effect of electron scattering (dotted curve) is very small, leading to a correction $\lesssim 1\%$ at $z\lesssim 1800$. Close to the maximum of the Thomson visibility function the effect of electron scattering is negligible. \change{This curve was computed using the numerical results in which we switched off electron scattering and then compared it to the one where it was included.}

Finally, we also computed the contribution that can be attributed to the effect of Doppler boosting and Doppler broadening (dash-dotted curve). For this we computed the escape probability when neglecting electron scattering and atomic recoil, but only including the line diffusion term. We then took the difference to result obtained in the no scattering approximation, as given earlier \citep{Chluba2008b}.
One can see that the diffusion term results in a decrease of the escape probability at low redshifts. However, this decrease is about 3 times smaller than the increase in the escape probability due to atomic recoil. Therefore the net effect due to resonance scattering is an increase in the escape probability, reaching $\Delta P/P \sim +4\%$ at $z\sim 600$.
\change{As explained in Sect.~\ref{sec:Pesc_def_num}, this shows that the decrease in the red wing escape probability due to the return of photons towards the Doppler core by line diffusion is more important than the increase of the escape fraction from within the Doppler core caused by Doppler broadening.}

\change{We would like to mention that the small variability in the diffusion contribution at $z\sim 600$ is likely due to some details in our numerical treatment. However, we expect that the corresponding result is converged at the $\sim 10\%$ level of the correction, which is sufficient for our purposes here.}

\subsection{Effect of the thermodynamic corrections factor}
\label{sec:fnu}
If we now in addition include the frequency-dependent asymmetry between the emission and absorption profile due to the thermodynamic correction factor which was introduced earlier \citep{Chluba2008b, Chluba2009}, we obtain the results presented in Fig.~\ref{fig:DP_fnu}. 
The dashed line again shows the correction in the no scattering approximation \citep{Chluba2009}.
The main correction do to the redistribution of photons over frequency again is due to the line recoil term (thin solid line). 
One can see that it is practically the same as in the previous case (see Fig.~\ref{fig:DP_phi_V}). Also the total correction due to electron scattering did not change very much.
\change{In both cases the difference was smaller than $\sim 5\%$ on the correction.}
However, the correction due to the line diffusion term seems to be slightly increased, suggesting a $f_\nu$ induced correction to the correction that is not completely negligible.
%

\begin{figure}
\centering
\includegraphics[width=0.9\columnwidth]{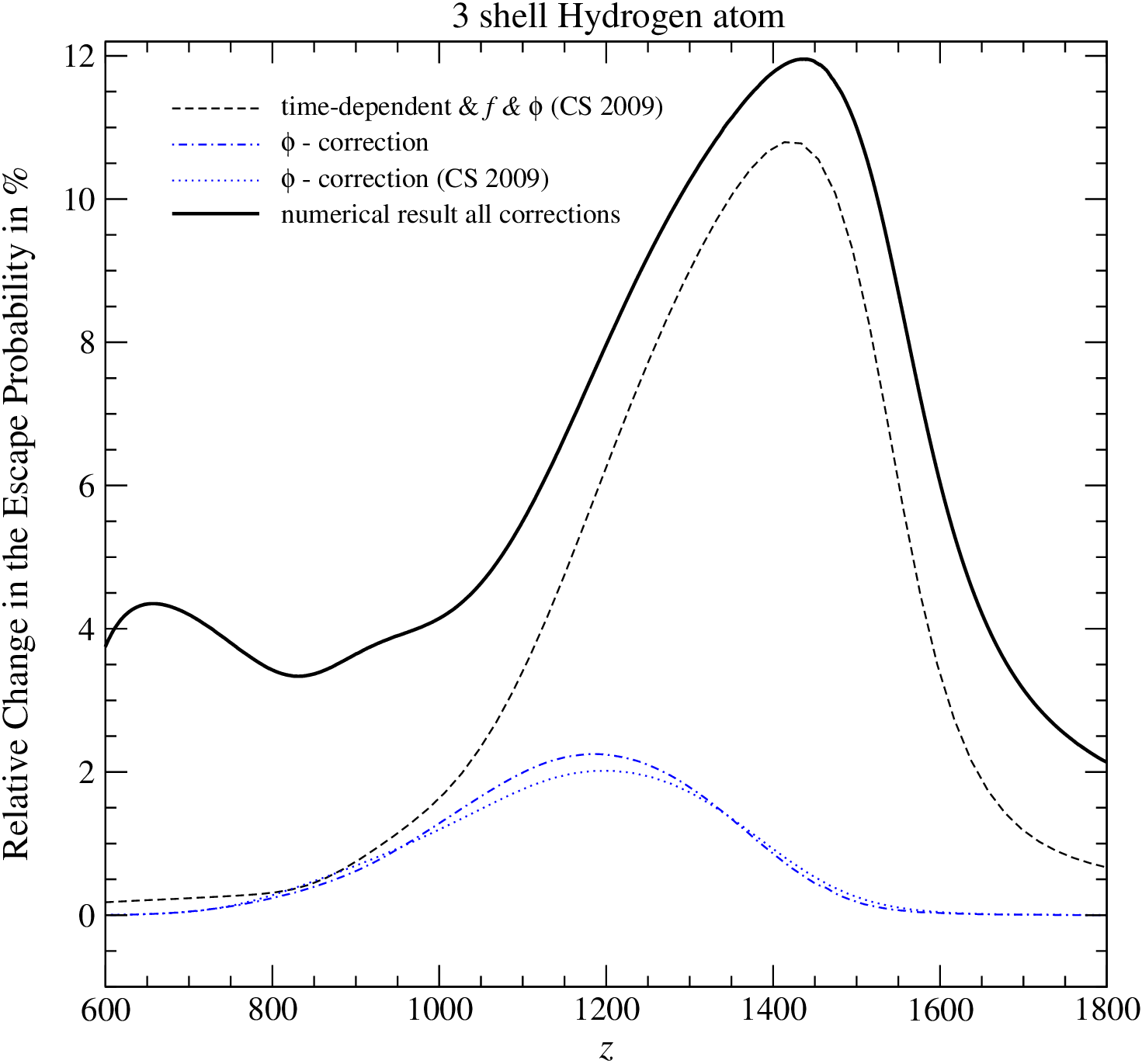}
\caption
{Changes in the Lyman $\alpha$ escape probability due to the shape of the emission profile. The dashed line shows the result obtained in the no scattering approximation \citep{Chluba2009}.}
\label{fig:DP_fphi}
\end{figure}
\subsection{Corrections due to the shape of the emission profile}
\label{sec:phi_2g}
Finally, we also ran the code including the correct shape of the 3d/3s-1s emission and absorption profile \citep{Chluba2009}. The results of these computations are shown in Fig.~\ref{fig:DP_fphi}. The dashed line again shows the correction in the no scattering approximation \citep{Chluba2009}. The dotted line in addition indicates the correction that was associated with the effect of the emission profile in the no scattering approximation \citep{Chluba2009}. 
We also computed the pure profile correction using the numerical results obtained when including the redistribution of photons and obtained the dash-dotted curve.  As one can see the difference to the no redistribution case is very small. Therefore we did not compute the pure recoil correction, the line diffusion correction or the correction due to electron scattering, since they should also be very similar to the contributions shown in Fig.~\ref{fig:DP_fnu}.

\section{Corrections to the ionization history}
\label{sec:changes_in_Xe}
In this Section we now give the expected correction to the ionization history when including the processes discussed in this paper.
For this we modified the {\sc Recfast} code \citep{SeagerRecfast1999}, so that we can load the pre-computed change in the Sobolev escape probability studied here.

\begin{figure}
\centering
\includegraphics[width=0.9\columnwidth]{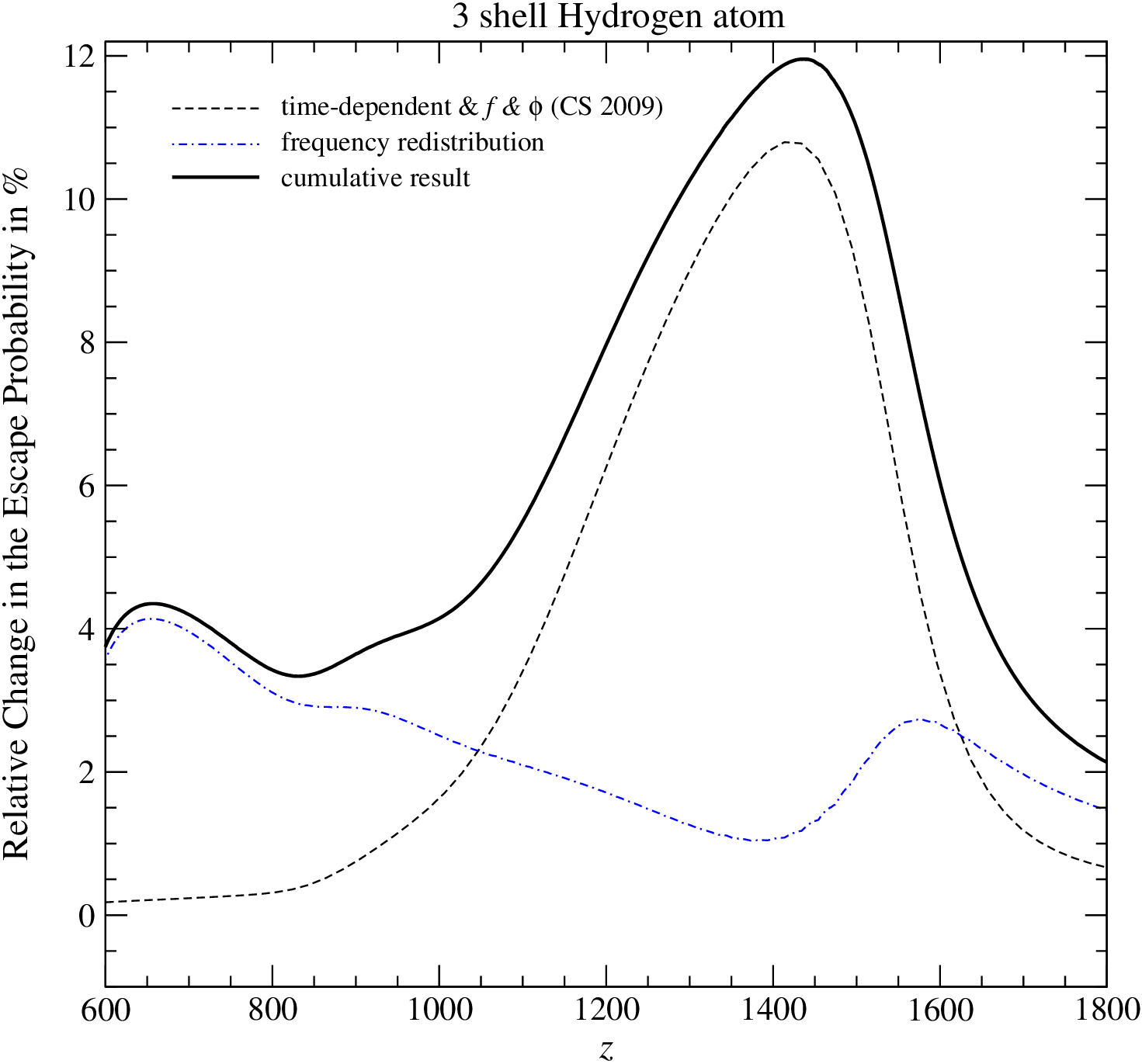}
\caption
{Changes in the Lyman $\alpha$ escape probability due to the different processes under discussion here. The dashed line shows the result obtained in the no scattering approximation \citep{Chluba2009}.}
\label{fig:DP_P.final}
\end{figure}

\begin{figure}
\centering
\includegraphics[width=0.9\columnwidth]{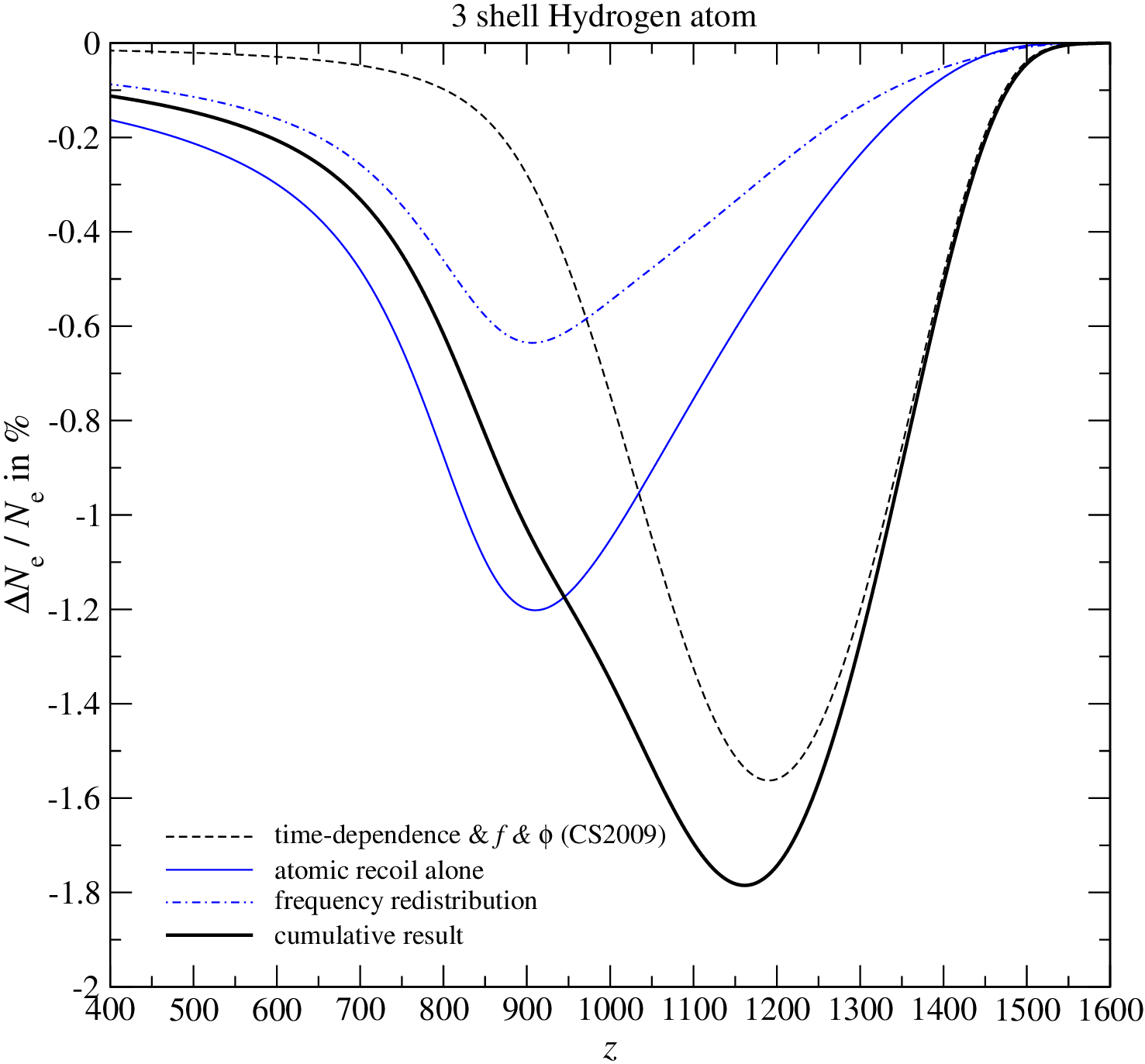}
\caption
{Changes in the free electron fraction due to the different processes under discussion here. The dashed line shows the result obtained in the no scattering approximation \citep{Chluba2009}.}
\label{fig:DN_N.final}
\end{figure}
%
In Fig.~\ref{fig:DP_P.final} we present the final curves for $\Delta P/P$ as obtained for the different processes discussed in this paper.
In Fig.~\ref{fig:DN_N.final} we show the corresponding correction in the free electron fraction computed with the modified version of {\sc Recfast}.
The atomic recoil effect alone (thin solid line) leads to $\Delta N_{\rm e}/N_{\rm e}\sim -1.2\%$ at $z\sim 900$. This is in very good agreement with the result of \citet{Grachev2008}. We already quoted this result earlier \citep[see footnote 10 in][]{Chluba2008b}, however there we just estimated the change in the free electron fraction using our full numerical result for the recoil correction on the Lyman $\alpha$ escape probability, without running it trough the {\sc Recfast} code.
Including electron scattering and all terms (line recoil and the diffusion term) for the redistribution of photons by the Lyman $\alpha$ resonance we obtain the dotted line. Here the total correction due to redistribution of photons now only reaches $\Delta N_{\rm e}/N_{\rm e}\sim -0.6\%$ at $z\sim 900$.
As we have seen in Sect.~\ref{sec:Lya_Pesc} this is due to the fact that the diffusion term slow recombination down again, since photons from the red wing return close to the Doppler core, where they die efficiently again.
%
%
Finally, the total correction including all the effects of photon redistribution and the correction due to the time-dependence, thermodynamic factor and shape of the profile, which were discussed earlier \citep{Chluba2009}, has a maximum of $\Delta N_{\rm e}/N_{\rm e}\sim -1.8\%$ at $z\sim 1160$.
Here the main contribution is coming from the the time-dependent correction and thermodynamic factor as explained in \citet{Chluba2009}.

\begin{figure}
\centering 
\includegraphics[width=0.95\columnwidth]{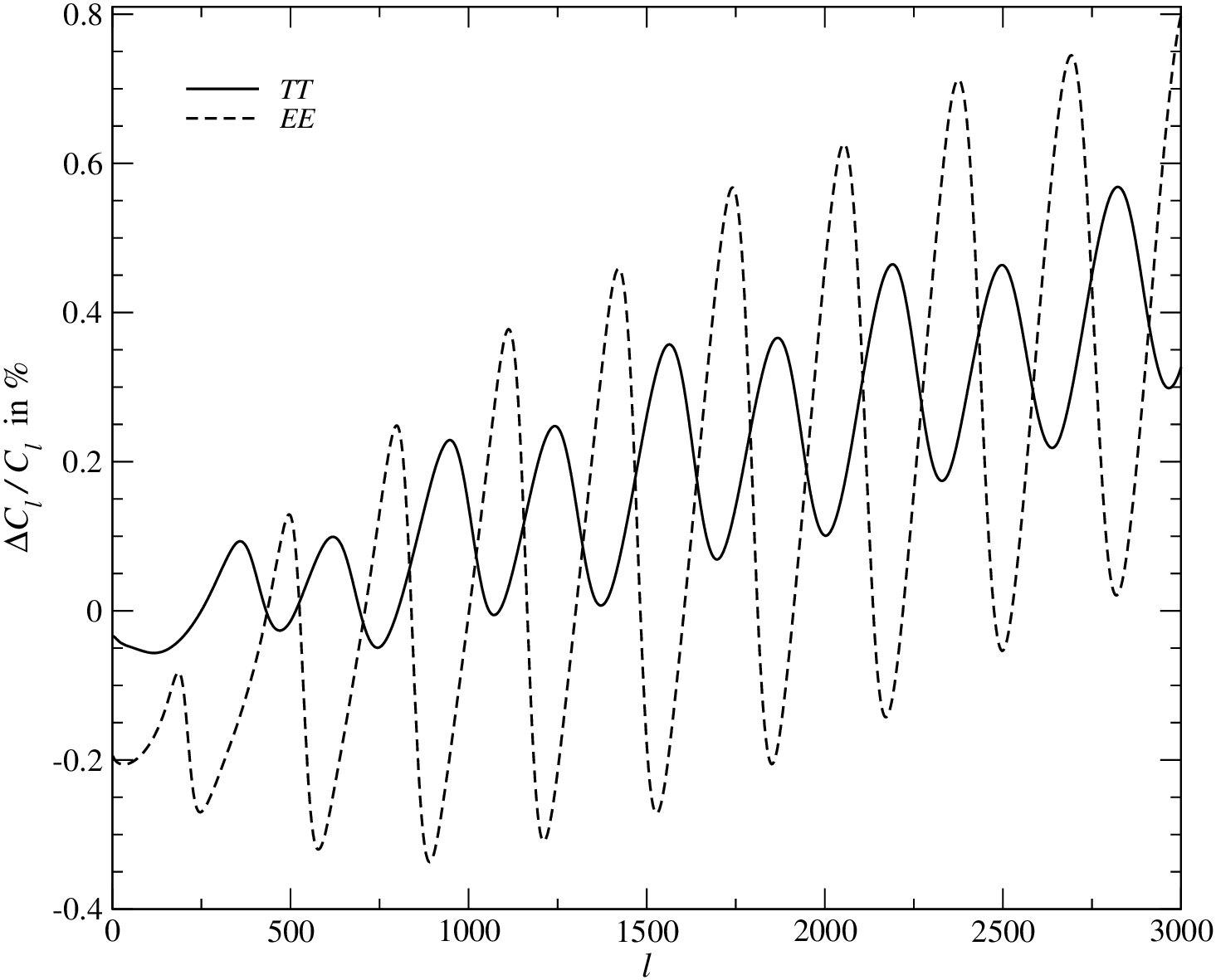}
\\[1mm]
\includegraphics[width=0.95\columnwidth]{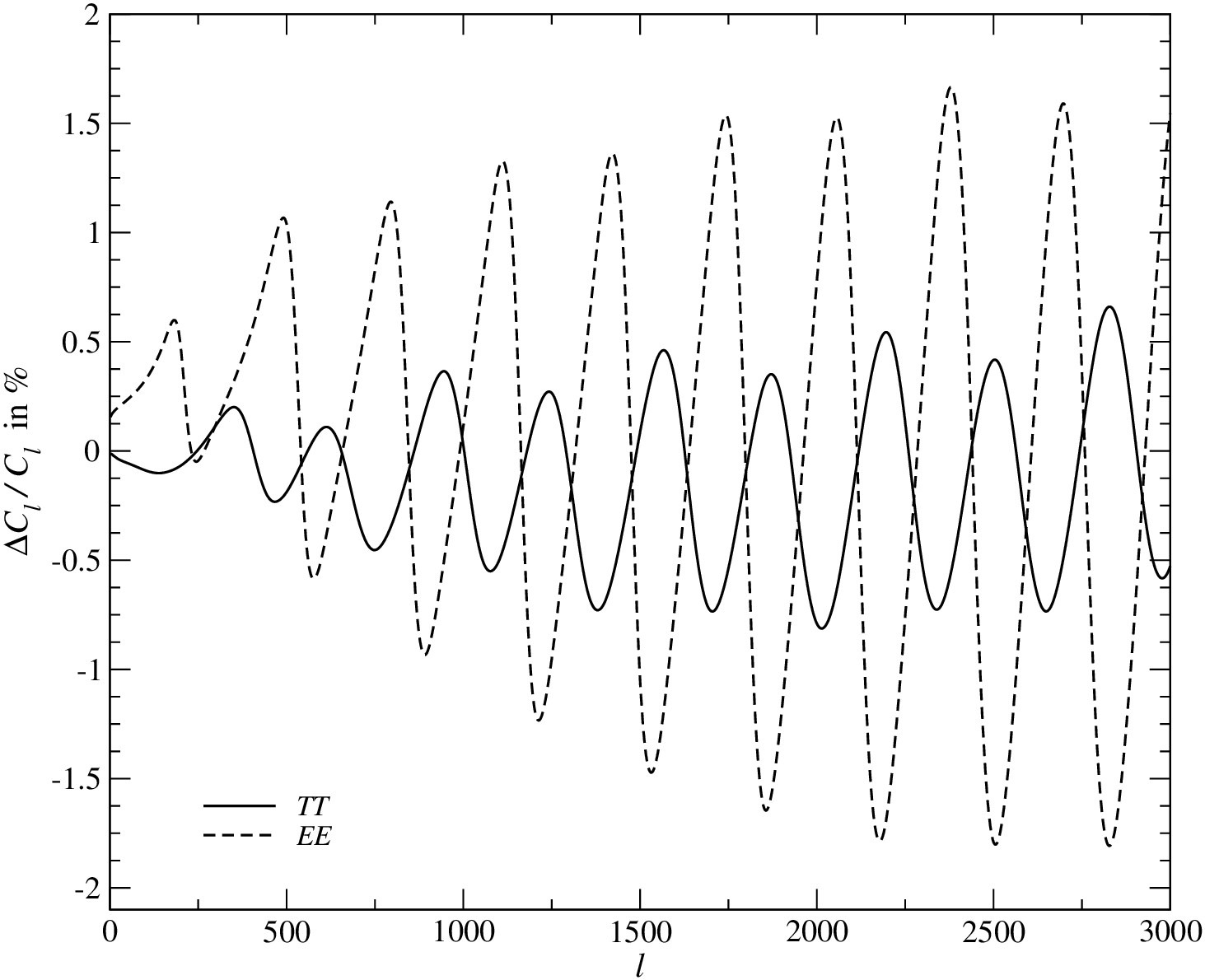}
\caption
{Changes in the CMB temperature and polarization power spectra. The upper panel shows the changes due to the redistribution of photons by line and electron scattering only. The lower panel shows the cumulative result in addition including the time-dependent correction, the thermodynamic factor, and the correction due to the shape of the emission profile, as discussed earlier \citep{Chluba2009}.}
\label{fig:DCl}
\end{figure}
In Fig.~\ref{fig:DCl} we finally show the changes in the CMB temperature and polarization power spectra.
The corrections to $\Delta N_{\rm e}/N_{\rm e}$ related to the redistribution of photons over frequency alone (upper panel) results in changes to the TT and EE power spectra, with peak to peak amplitude $\sim 0.5\%-1\%$ at $l\gtrsim 1500$.
When also including the processes discussed in \citet{Chluba2009} at $l\gtrsim 1500$ we find a cumulative correction of  $|\Delta C_l/C_l|\sim 1\%$ for the TT power spectrum and $|\Delta C_l/C_l|\sim 2\%-3\%$ for the EE power spectrum.
It will be important to take these changes into account in the analysis of future CMB data.

\section{Conclusions}
\label{sec:disc_con}
In this paper we have considered the effect of {\it frequency redistribution} on the escape of Lyman $\alpha$ photons during hydrogen recombination.
We have shown that line recoil speeds hydrogen recombination up by $\Delta N_{\rm e}/N_{\rm e}\sim -1.2\%$ at $z\sim 900$.
On the other hand, the combined effect of Doppler boosting and Doppler broadening at different distances from the line center slows hydrogen recombination down by $\Delta N_{\rm e}/N_{\rm e}\sim +0.6\%$ at $z\sim 900$. 
As explained in Sect.~\ref{sec:illustr}, line diffusion (including both Doppler boosting and Doppler broadening) increases the escape fraction for photons that are emitted in the vicinity of the Doppler core in comparison with the value obtained in the no scattering approximation. In particular some small fraction of photons that are emitted on the blue side of the resonance can still escape, since due to line diffusion they pass through the Doppler core faster than dying there.
On the other hand, for photons that are emitted at $-\text{few}\times 10^2 \lesssim \xD\lesssim -10$ (i.e. in the red wing) it becomes harder to escape, since line diffusion brings some of these photons back close to the Doppler core, where they are absorbed efficiently. 
For photons that are emitted at  $\xD\lesssim -\text{few}\times 10^2$ the redistribution over frequency can be neglected.
We also showed that electron scattering has a minor effect on recombination dynamics at redshifts $z\lesssim 1400$.
In total the redistribution of photons over frequency leads to a speed up of hydrogen recombination by  $\Delta N_{\rm e}/N_{\rm e}\sim -0.6\%$ at $z\sim 900$ (cf. Fig.~\ref{fig:DN_N.final}).
This results in important changes to the CMB temperature and polarization power spectra (see Fig.~\ref{fig:DCl} for details), which should be taken into account for the analysis of future CMB data.

In addition, we would like to mention that the cumulative changes (including the processes discussed in \citet{Chluba2009} and those of this work) in the Lyman $\alpha$ photon escape probability will be very important for precise computations of the cosmological recombination spectrum \citep[e.g. see][for review and references]{Sunyaev2007}. Here it is interesting that the changes in the shape of the recombination lines connected with electrons passing through the Lyman $\alpha$ channel are expected to be $\sim 10\%$ at $z\sim 1400$ (in comparison to $\sim 2\%$ for $N_{\rm e}$ at $z\sim 1200$). Observing the cosmological recombination lines and looking at their exact shape would therefore provide a more direct and $\sim 4-5$ times more sensitive probe for the physics of cosmological recombination than with the CMB temperature anisotropies.

\begin{appendix}

\section{Computational details}
\label{app:comp_details}

\subsection{Solver for the differential equations}
\label{app:Solver}
In order to solve the photon transfer equation we used the solver {\sf D03PPF}
from the {\sc Nag}\footnote{www.nag.co.uk}-Library. It provides possibilities
for extensive error control and adaptive remeshing. In particular for
computations with narrow initial spectra or low line scattering efficiency
this feature became very important. However, remeshing also leads to an
additional loss of accuracy for long integrations and therefore has to be
applied with caution.

Typically we used $\sim 2500-5000$ grid-points for the representation of the
photon distribution and required relative accuracies $\epsilon\sim
10^{-6}-10^{-5}$.
We checked the convergence of the results by varying the accuracy requirements
and number of grid-points, and also by running several test problems for which
analytic solutions exist.

\end{appendix}


\bibliographystyle{aa} 
\bibliography{Lit}

\end{document}